\documentclass[sigconf,screen]{acmart}

\AtBeginDocument{%
  }

\copyrightyear{2026}
\acmYear{2026}
\setcopyright{cc}
\setcctype{by}
\acmConference[ASE '26]{Proceedings of the 41st IEEE/ACM International Conference on Automated Software Engineering}{October 12--16, 2026}{Munich, Germany}
\acmBooktitle{Proceedings of the 41st IEEE/ACM International Conference on Automated Software Engineering (ASE '26), October 12--16, 2026, Munich, Germany}
\acmDOI{10.1145/3832783.3834441}
\acmISBN{979-8-4007-2882-2/2026/10}

\usepackage{multirow}
\usepackage{makecell}
\usepackage[linesnumbered,ruled,vlined]{algorithm2e}
\usepackage{graphicx}
\usepackage{subcaption}

\usepackage{booktabs}
\usepackage{enumitem}
\usepackage{soul}

\usepackage[]{collab}

\usepackage{xspace}
\def\tool{\textit{MicroAgent}\xspace}

\begin{document}

\title{MicroAgent: Context-Augmented Multi-agent Framework for Automatic Microservice Decomposition}

% \affiliation{%
\author{Zishan Su}
\email{zshsu@link.cuhk.edu.hk}
% \authornotemark[2]
\affiliation{%
  \institution{The Chinese University of Hong Kong}
  \city{Hong Kong}
  \country{China}
}

\author{Junjie Huang}
\email{jjhuang23@cse.cuhk.edu.hk}
% \authornotemark[2]
\affiliation{%
  \institution{The Chinese University of Hong Kong}
  \city{Hong Kong}
  \country{China}
}

\author{Shiwen Shan}
\email{shanshw@mail2.sysu.edu.cn}
% \authornotemark[3]
\affiliation{%
  \institution{Sun Yat-sen University}
  \city{Zhuhai}
  \country{China}
}

\author{Xingyan Chen}
\email{chenxy979@mail2.sysu.edu.cn}
% \authornotemark[3]
\affiliation{%
  \institution{Sun Yat-sen University}
  \city{Zhuhai}
  \country{China}
}
\author{Hui Zeng}
\email{zengh76@mail2.sysu.edu.cn}
% \authornotemark[3]
\affiliation{%
  \institution{Sun Yat-sen University}
  \city{Zhuhai}
  \country{China}
}
\author{Yuxin Su}
\authornote{Yuxin Su is the corresponding author.}
\email{suyx35@mail.sysu.edu.cn}
% \authornotemark[3]
\affiliation{%
  \institution{Sun Yat-sen University}
  \city{Zhuhai}
  \country{China}
}
\author{Yanlin Wang}
\email{wangylin36@mail.sysu.edu.cn}
% \authornotemark[3]
\affiliation{%
  \institution{Sun Yat-sen University}
  \city{Zhuhai}
  \country{China}
}
\author{Michael R. Lyu}
\email{lyu@cse.cuhk.edu.hk}
% \authornotemark[2]
\affiliation{%
  \institution{The Chinese University of Hong Kong}
  \city{Hong Kong}
  \country{China}
}


\begin{abstract}
The adoption of Microservice Architecture (MSA) has revolutionized software engineering by enhancing scalability, agility, and maintainability over traditional monolithic applications. 
As more developers transition their legacy systems to microservice-based architectures, effective microservice decomposition—partitioning monolithic applications into highly cohesive services—becomes vital.
% due to its profound influence on system performance and sustainability.
However, this decomposition task presents significant challenges. Manual approaches are time-consuming and labor-intensive. Existing automated methods often fail to capture the necessary semantic insights from complex applications, while naive applications of Large Language Models tend to overlook crucial contextual information and design principles, leading to suboptimal results.

To address these challenges, we propose \tool, a Context-Augmented Multi-Agent Framework for Microservice Decomposition. 
Our framework divides the decomposition process into five distinct subtasks and assigns each to a specialized agent. 
To enhance the effectiveness of each agent, we provide tailored, multi-granularity context that keeps its analysis focused and mitigates information overload.
Furthermore, to ensure the decomposition adheres to established design principles, we integrate analytical tools that guide the agents' decision-making. 
Experimental evaluations on 10 Java Web applications demonstrate that \tool achieves an average decomposition accuracy of 89.2\%, outperforming the state-of-the-art method by 24.6\%.
We also conduct a case study to highlight the practical benefits of our design.
\end{abstract}


%    <concept>
%        <concept_id>10011007.10011074</concept_id>
%        <concept_desc>Software and its engineering~Software creation and management</concept_desc>
%        <concept_significance>500</concept_significance>
%        </concept>
%  </ccs2012>

% \ccsdesc[500]{Software and its engineering~Software creation and management}




% \setcopyright{none}
% \settopmatter{printacmref=false}
% \renewcommand\footnotetextcopyrightpermission[1]{}
\maketitle

\section{Introduction}
\label{sec:introduction}
% 1. 微服务是XXX
% 2. 相比于monolithic，微服务有什么优点
% 3. 因此越来越多公司将应用迁移诚微服务架构
Microservice Architecture (MSA) has become a popular paradigm for building complex systems as a composition of small, independent components that communicate with each other through network protocols such as HTTP~\cite{lewis2014microservices,thones2015microservices}.
Compared to monolithic architectures that are packaged as one complex deployment unit, microservices provide improved scalability, agility, and maintainability~\cite{quattrocchi2024cromlech,wang2024microservice}. 
Consequently, many companies, such as Amazon and GitHub~\cite{III_2021,Uchitelle_2021}, have migrated their legacy monolithic applications to microservice-based solutions.

% %大背景
% As monolithic applications grow increasingly bulky and complex, their tightly coupled nature gives rise to inflexibility, which makes them difficult to maintain and challenging to scale for developers.
% % yx: 单体程序是什么，这个得事先说明。另外，微服务是云计算的产物
% Consequently, microservice architecture has emerged as an attractive alternative. 
% It decomposes applications into smaller, independent, highly cohesive, and loosely coupled services, each typically focused on a specific functionality. 
% This approach facilitates independent development, as well as elastic scaling and improved resource utilization \cite{?}. 
% Due to these benefits, many developers and companies, such as GitHub and IBM, are eager to migrate their monolithic applications to a microservice architecture \cite{?}. 

% 1. 微服务分解是迁移中重要的第一步，他是XXX。因为它影响后续开发流程【TSC-24 Cromlech】
% 2. 然而，manual decomposition耗时费力，因为complex code dependency and diverse requirements【TSC-24 Cromlech】.
The very first task in this migration is \emph{microservice decomposition}, which partitions the existing monolith into a set of highly cohesive and loosely coupled services~\cite{newman2021building,richardson2018microservices,abgaz2023decomposition}. 
The decomposition results can significantly affect the maintainability, scalability, and performance of the resulting system, since HTTP communication among MSA has a significantly higher overhead compared to local function calls in monolithic applications~\cite{quattrocchi2024cromlech}.
% \orange{This design principle is particularly important because HTTP communication has a significantly higher overhead compared to local function calls in monolithic applications.}
% Moreover, effective decomposition is essential, as it directly impacts the maintainability, scalability, and performance of the resulting system~\cite{TSC-24 Cromlech}. 
Traditionally, this task heavily relies on manual analysis of codebases and business logic~\cite{carvalho2019analysis,taibi2018definition}, making it time-consuming and labor-intensive, especially for large and complex legacy systems with intricate dependencies~\cite{quattrocchi2024cromlech}. 
To address these challenges, various automatic decomposition methods have been proposed.

% %小背景
% However, for existing legacy monolithic applications, manually converting them into microservices is labor-intensive and time-consuming. 
% % why convert them labor-intensive and time-consuming 
% The very first step of migration is to decompose the monolithic application and identify the microservice candidates. 
% Hence, in both academia and industry, automating the decomposition of monolithic applications into microservices has become a topic of wide interest.

% Existing decomposition methods typically transform monolithic applications into a relational graph by extracting relevant information.
% % and extract relevant information to build a relation graph.
% Then, clustering algorithms are applied to partition the application into microservice candidates.

Existing methods often take a two-step framework for decomposition, which first extracts relevant features from the monolithic application and then applies clustering algorithms to aggregate cohesive functional units for each microservice based on these features~\cite{wang2024microservice,nitin2022cargo}.
% Specifically, a graph model is 
The information is encoded into a graph, where nodes symbolize application elements (e.g, classes, methods), and edges represent certain relationships between them.
% Then, a clustering algorithm is employed to partition the application into microservice candidates, with the optimization goal of forming cohesive clusters.
To enhance subsequent clustering, various methods have been proposed for robust feature extraction, which can be categorized into two types: \textit{program analysis-based}~\cite{nitin2022cargo,kalia2020mono2micro,assunccao2021multi,filippone2023monolithic,liu2022log2ms} and \textit{semantic analysis-based methods}~\cite{sellami2022combining,mazlami2017extraction,al2021microservice,trabelsi2023legacy}. 
% The effects of the methods are primarily dependent on the heuristic they adopt to extract graph information, which usually involves \textit{program analysis}\cite{} and \textit{semantic analysis}\cite{}.

% \jj{use semantic analysis rather than linguistic analysis. Linguistic is inaccurate. Then this para and next para should be one}
% Then by utilizing clustering algorithms or community detection methods, the graph is partitioned into microservice partitions.
% According to the heuristic they adopt, the approaches usually fall into different categories.

\textit{Program analysis-based methods} utilize static or dynamic analysis to extract structural information (e.g. call graph relations, control flow relations) of the monolithic applications to build corresponding graphs.
% the intrinsic limitations of static analysis, these approaches often suffer from low coverage, leading to inaccurate and incomplete decomposition results for larger-scale cases.
% Moreover, the tools 
% existing static analysis-based methods\cite{} are often imprecise, and they may rely on tools tailored to specific frameworks, lacking support for popular web frameworks (e.g. Spring Boot)\cite{}.
% This limits their generalizability to diverse architectures and large-scale legacy codebases~\cite{cite for problem of large scale}.
% However, static analysis-based methods\cite{} inherently struggle with precision in large-scale projects\cite{}, proving to be difficult to fully capture call relationships in the monolithic applications.
However, static analysis-based methods~\cite{nitin2022cargo,filippone2023monolithic} often struggle to reconstruct comprehensive service invocation relations as critical information, such as routing and dependency, is distributed across configurations beyond static source code. Consequently, they cannot overcome the fragmentation and dynamic nature intrinsic to the original monolith applications.
%monolithic architectures.
% However, static analysis-based methods inherently struggle with precision in large-scale projects\cite{}, proving to be difficult to fully capture call relationships in the monolithic applications.
Dynamic analysis-based methods~\cite{kalia2020mono2micro,liu2022log2ms}, on the other hand, often employ use-case-based analysis with extensive test cases to achieve sufficient code coverage; yet, most real-world monoliths lack such coverage, leading to incomplete dependency extraction and suboptimal decomposition.
To complement the sparse information obtained from program analysis, some works~\cite{sellami2022combining,al2021microservice,mazlami2017extraction} attempt to incorporate \textit{semantic analysis} into graph construction for clustering.
These works mostly use off-the-shelf techniques (e.g. Term Frequency, embedding models) to obtain representations for the code elements,
which only extract local and shallow semantic features such as class/method names similarity,
lacking a global understanding of the in-depth business logic.
% codebase.
% hindering the contextual information of the whole application and the expressiveness of business logic the application contains.
% contextual information in code
% failing to capture microservice-specific contextual information.
% Another category employs neural networks to learn linguistic representations for monolithic applications.
% To enhance the expressiveness of the embeddings, these works\cite{} either train on certain applications, or fine-tune based on a microservice-related dataset, which requires abundant data collection.
As a result, existing solutions often fail to achieve accurate and scalable microservice decompositions.

The advent of Large Language Models (LLMs) opens up new possibilities for addressing the microservice decomposition task. 
% \red{Having been trained on vast datasets, LLMs have acquired abundant knowledge in both code understanding and microservice architecture.
% Moreover, they have demonstrated strong capabilities in completing tasks when provided with proper instructions. }\jj{the motivation of using LLM should align with the limitation of program analysis-based and semantic-based methods. Try to explain more about why LLM-based methods can address the limitations in a high level.}
As LLMs have been trained on large-scale corpus in diverse domains (e.g., code and business), they can leverage the comprehensive pre-trained knowledge to understand existing codebases and business logic that the applications imply~\cite{fan2023large}.
However, directly leveraging LLMs to microservice decomposition presents the following challenges: 
% Firstly, the limited context window of LLMs restricts the amount of code and related information they can process at once~\cite{llm context length}, which often prevents them from capturing the full system structure and leads to fragmented or incoherent service boundaries. 
% Secondly, LLMs tend to struggle with complex inferring complex relationships such as class dependencies and cross-module interactions in large codebases~\cite{cite some repo code gen work}, resulting in tightly-coupled classes being split across services and thus increasing maintenance complexity. 
% In addition, LLMs may not fully parameterize key microservice design principles, such as low coupling and high cohesion~\cite{cite microservice priciple}, which can cause duplication of domain-specific classes or improper allocation of common utilities, ultimately leading to excessive inter-service communication.
(1) \textbf{Overlong context length}: LLMs can be overwhelmed with the repository-level application codebase, as their performance deteriorates when the number of input tokens increases \cite{li2024long,li2023loogle}. This limitation can hinder the complete analysis of the codebase, resulting in fragmented or incoherent microservice boundaries. 
% that hinder effective service isolation and scalability.
(2) \textbf{Lacking context insight:} Due to the complex contextual information the applications express, LLMs may struggle to develop a deep-going insight into concrete context, such as class dependencies in complicated applications~\cite{zhang2025llm}.
% For instance, they may fail to identify class dependencies in complicated applications\cite{} 
This can lead to incorrect decomposition, where interdependent classes are improperly segregated, causing maintenance challenges in the resulting microservices.
(3) \textbf{Oversight on domain knowledge:} LLMs can overlook core microservice principles and produce impractical decompositions.
For example, they may allocate domain-specific classes into irrelevant services, inducing high coupling; or they may misassign common utilities to a single service, triggering excessive inter-service communication in practical systems.

% \st{(1) Performance Degradation with Repository-Level Code Context: Monolithic application code exists at the repository level, making it impractical to input directly into an LLM.}
% \st{Even if the input does not exceed the token limit, studies have shown that task performance deteriorates significantly as the number of input tokens increases.}
% \st{Therefore, when faced with raw monolith code, LLMs struggle to gain a systematic understanding of the application, often omitting details or generating hallucinations.}
% \st{(2) Over-Reliance on Semantic Similarities of Code Elements: While LLMs excel at understanding semantic information, they may overly rely on semantic similarities during microservice decomposition, overlooking code dependencies.}
% \st{For instance, LLMs might group classes with similar root words together, even when code elements relationships indicate they belong to different microservices.}
% \st{(3) Imprecise Identification and Assignment of Common Classes: For the problem of identifying and assigning common classes, which is not addressed by prior work, LLMs also face challenges.}
% \st{They may fail to capture nuanced call relationships, mistakenly assigning common classes to microservice candidates; or simply distribute common classes to one of the microservice candidates, which is considered as an anti-pattern in microservice design. Disposing common classes improperly can result in high delays in cross-service communication, which is a severe problem in real-world implementations.}

% 因此我们考虑用agent来解决以上问题
% \textbf{Our work.}
To address these issues, we propose \textbf{\tool}, a Context-Augmented Multi-Agent Framework for Microservice Decomposition. 
Specifically, we divide the microservice decomposition task into five subtasks, each handled by specific agent(s), including Domain Agent, Clustering Agents, Merging Agent, Common Class Agent, and Review Agent.
By breaking down the overall task, each agent can focus on a specific scope of the task, without concentrating on all the repository-level information,
thus tackling the first challenge of overlong context.
While decomposing the original task into several stages inherently narrows down the context each agent needs to consider, this brings about another problem of what exact context should be exposed to each agent~\cite{langchainContextEngineering,llamaindexContextEngineering}.
Hence, we extract and compress the original codebase at varying levels of granularity to generate customized context tailored to each stage in microservice decomposition, covering both application-level and class-level. 
These contexts are adaptively provided to the agents based on the specific subtask scenario, ensuring each agent has access to the most task-relevant information to delve into.
This resolves the second challenge of lacking context insight.
To further empower agents with contextual awareness of decomposition tasks and align them with core microservice design principles, we also introduce a suite of specialized tools. 
This toolkit provides additional decomposition-oriented feedback for agents to assist in their dynamic adjustment,
addressing the third challenge of oversight on domain knowledge.
% \red{This framework facilitates more accurate and sound decomposition results by providing agents with bespoke context and specialized toolkit.}\jj{don't say accurate results. Just introduce methods in high level.} \jj{No need to use three bullet points here to introduce to much details in each part. Just show overall idea and novel point.}

% obtaining quantitative statistics, helping them efficiently identify task breakthroughs. 
% For example, in aggregation tasks, we provide semantic similarity search tools and dependency retrieval tools to identify classes related to a domain. 
% For common tasks, we develop entropy-related metrics to discover potential common classes. 
% Our overall strategy in tool design is to avoid agents exploring aimlessly among numerous classes, reducing cognitive complexity, while also allowing tools to assist rather than dominate their decision-making.

% evaluation
We evaluate \tool based on a benchmark of monolithic applications with corresponding microservice versions.
Our results show that \tool achieves an 89.2\% accuracy in microservice decomposition on average, which notably surpasses the best baseline --  the LLM base model by 24.6\%.
Particularly, to demonstrate that our framework has incorporated core microservice design principles and can generate practical decomposition results, we measure the effects of common classes identification and assignment, which are often neglected in previous works.
In terms of this, \tool showcases a 93.4\% F1 score, outperforming the state-of-the-art by 41.1\%.

% regarding the effects of common classes identification and assignment, our approach demonstrates a 100\% recall rate and 97.8\% F1 score, outperforming the best baseline by 208.6\% and 108.5\% respectively.

% contribution
% \textbf{Contributions.}
To sum up, the main contributions of this work are as follows:

\begin{itemize}[leftmargin=*]
    \item To the best of our knowledge, we propose the first LLM-based agentic framework for microservice decomposition, 
which decomposes monolithic applications into cohesive microservice partitions, with five agents for tailored subtasks.
    \item We enhance the agents' contextual understanding through a dual strategy: providing customized, multi-granularity contexts tailored to each subtask, and equipping them with decomposition-oriented analytical tools. This approach ensures that agents focus on the most relevant information, leading to practical and accurate microservice partitions.
    \item We evaluate \tool on the benchmark including 10 Java web applications. The results show that \tool achieves an accuracy of 89.2\% in microservice decomposition, outperforming the state-of-the-art method by 24.6\%.
Regarding common class identification and assignment,
\tool attains a 93.4\% F1 score, improving the best baseline by 41.1\%.
We also conduct a case study to illustrate how \tool produces practical decomposition.
\end{itemize}

\section{Background and Motivation}
\label{sec:background_motivation}

\subsection{Microservice Decomposition}

Microservice architecture is a typical software design paradigm where applications are structured as a collection of small, loosely-coupled services, each responsible for a distinct aspect of the business logic~\cite{lewis2014microservices}. 
% These systems usually consist of multiple core components: \textit{domain-specific code} that handles specific business functionalities and manages their own data; \textit{common code} that provides reusable features such as utility classes or public interfaces;
% % authentication and data models
%  \textit{API gateways} that act as unified entry points for client requests and handle routing and authentication; \textit{infrastructure modules} including databases, message queues, and service registries; and \textit{communication mechanisms} such as RESTful APIs, gRPC, or asynchronous message queues that enable interaction between services. 
In practice, a microservice system typically contains: 
(i) \textit{domain-specific code} that implements specific business logic (i.e., an application domain such as ordering, payment, or shipping) and manages domain data; 
(ii) \textit{common code} that provides reusable capabilities or shared artifacts (e.g., utility classes, shared interfaces, or data contracts); 
(iii) \textit{API gateways} that serve as unified entry points for client requests and handle routing and authentication; 
(iv) \textit{infrastructure components} such as databases, message queues, and service registries; 
and (v) \textit{communication mechanisms} (e.g., REST, gRPC, or asynchronous messaging) that enable inter-service interaction.
% However, the sheer size and diversity of these components introduce significant complexity to manual development and maintenance. To address these challenges, frameworks such as Spring Cloud~\cite{spring} and Netflix OSS~\cite{netflix oss} have been developed to simplify the management of microservice systems.

To migrate from a monolithic application and form a microservice system, the first step is to decompose the complex monolith into a set of disentangled microservice candidates~\cite{abgaz2023decomposition}. 
This process is guided by a Domain-Driven Design (DDD)~\cite{evans2004domain} principle, which has been widely adopted in microservice design, such as Microsoft~\cite{microsoftDomainMicroservice}. 
DDD focuses on aligning software structure with application domains, encouraging developers to encapsulate related business logic and data within clear boundaries.
By doing so, DDD helps ensure that each microservice corresponds to a specific business capability and remains cohesive and maintainable. 
In this work, we follow most previous works and adopt DDD as the primary design principle for microservice decomposition. Specifically, we combine codebase and database analysis with domain-knowledge comprehension to automatically identify and map application domains within the monolithic application, facilitating the transformation into cohesive, domain-aligned microservices.

\subsection{Motivating Examples}

\begin{figure}[htbp]
    \centering
    \begin{subfigure}[b]{\linewidth}
        \centering
        \includegraphics[width=0.9\textwidth]{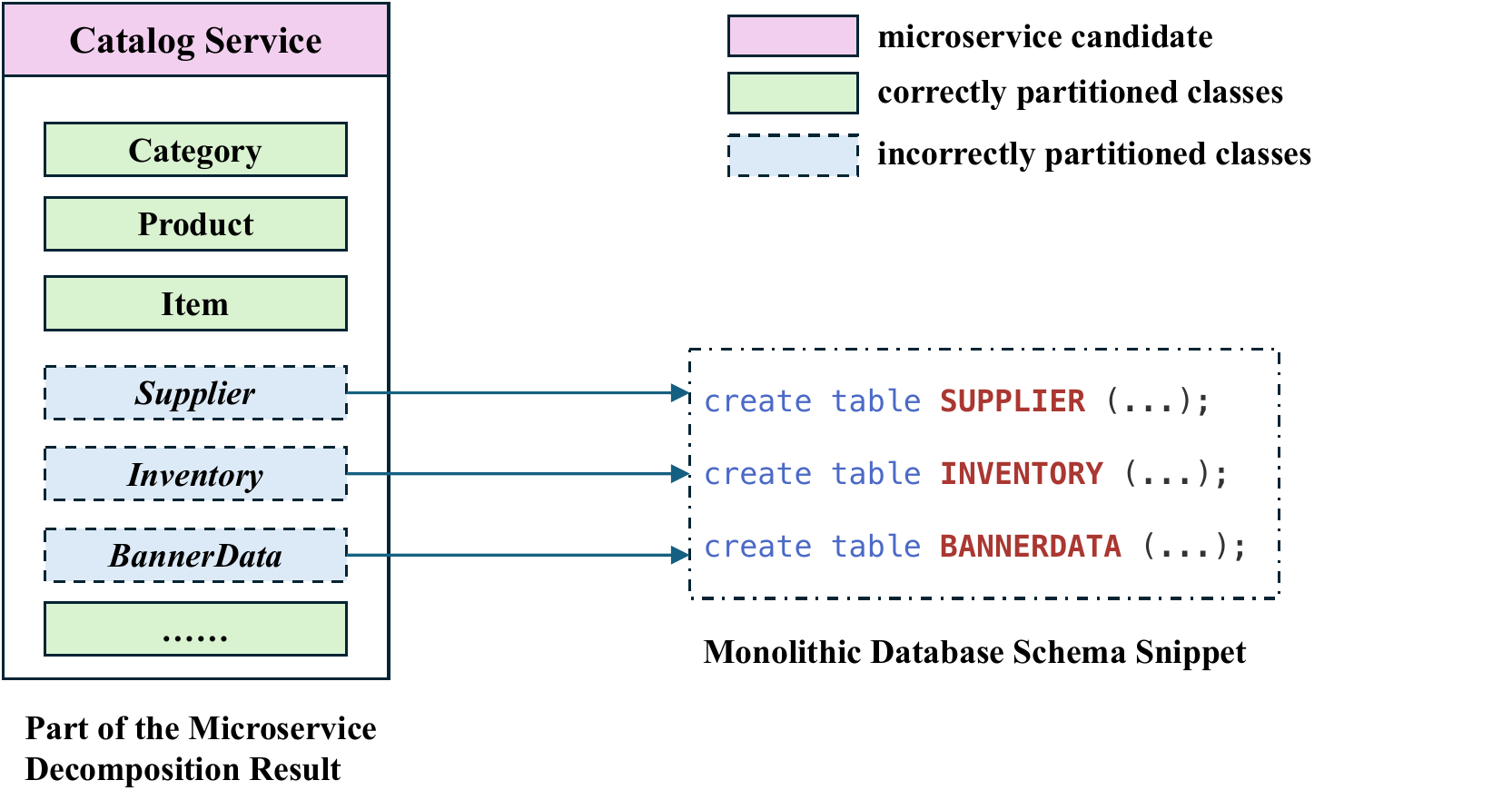}
        \caption{
        % A demonstration example of \textit{Catalog Service} in the microservice decomposition of JPetstore generated by LLM.
        Example of LLM hallucination when provided with repository-level context: non-existent classes ``\textit{Supplier}'', ``\textit{Inventory}'', and ``\textit{BannerData}'' are assigned to the \textbf{Catalog Service} partition. }
        \label{fig:exp1}
    \end{subfigure}

    \bigskip
    
    \begin{subfigure}[b]{\linewidth}
        \centering
        \includegraphics[width=0.9\textwidth]{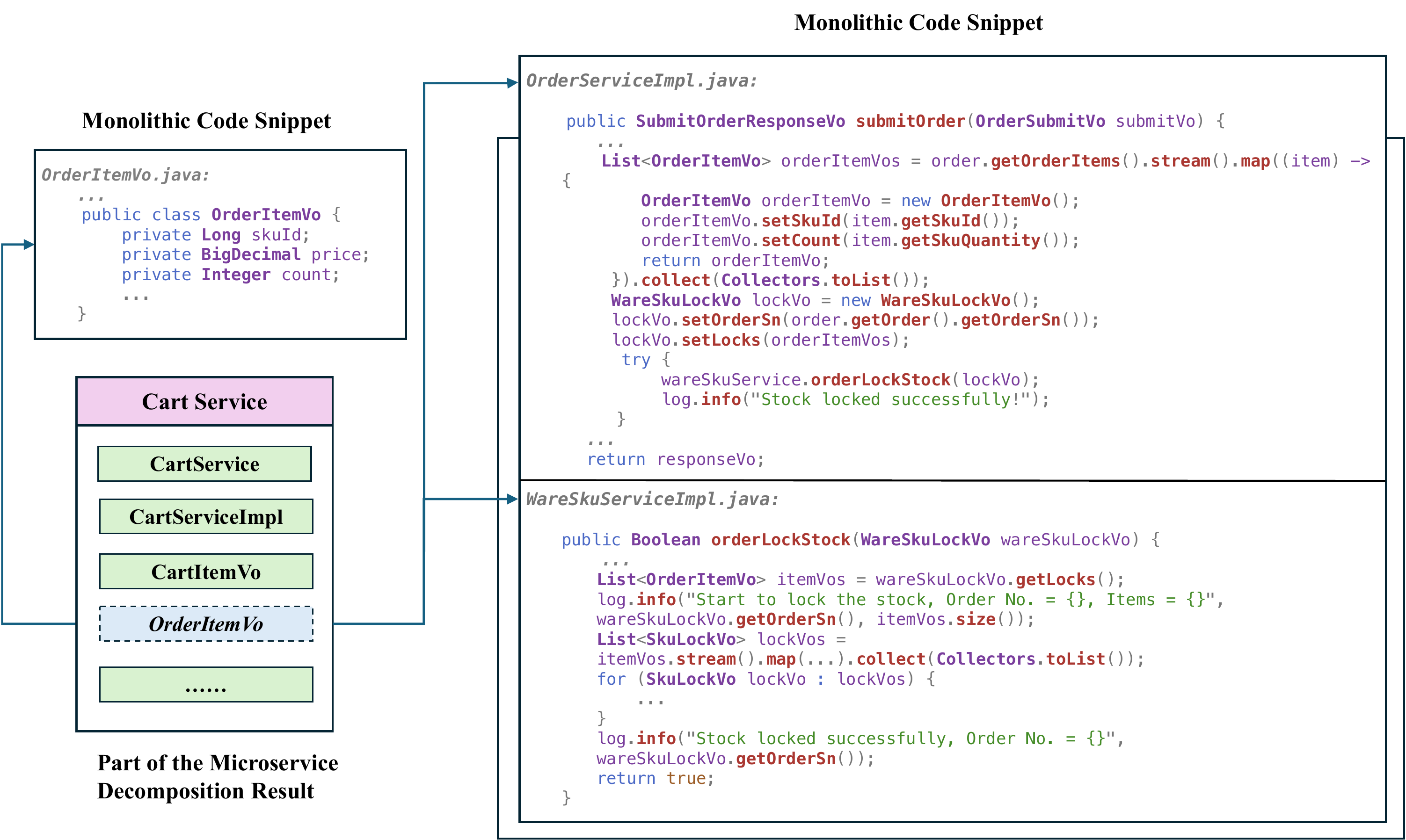}
        \caption{
        % A demonstration example of \textit{Cart Service} in the microservice decomposition of guli-mall generated by LLM.
        Example of LLM's insufficient use of concrete code context: 
        ``\textit{OrderItemVo}'' is assigned to \textbf{Cart Service}, while the code snippet shows it is used in order submission and stock-locking logic.
        }
        \label{fig:exp2}
    \end{subfigure}
    
    \caption{Motivating examples demonstrating the challenges of directly utilizing LLMs for microservice decomposition.}
    \label{fig:example}
    % \vspace{-15pt}
\end{figure}

In this section, we use two illustrative real-world examples to demonstrate the challenges of directly applying LLMs to decompose the monolithic version of JPetstore~\cite{githubGitHubMybatisjpetstore6} and gulimall~\cite{gulimall}.
JPetstore~\cite{githubGitHubMybatisjpetstore6} is a pet-store web application, and gulimall~\cite{gulimall} is an online shopping mall application.
We provide Deepseek-V3.2~\cite{liu2025deepseek} and GPT5.2~\cite{openaiIntroducingGPT52} with the monolith's raw source code and the database file, and require LLMs to output the class-level microservice decomposition result.

Figure~\ref{fig:exp1} shows the LLM-generated decomposition for \textbf{Catalog Service} in JPetstore application, in which classes named ``\textit{Supplier}'', ``\textit{Inventory}'' and ``\textit{BannerData}'' are assigned to this service partition.
However, these three terms only appear as tables in the database schema and do not correspond to any classes in the source code, suggesting that LLM may mistakenly interpret database entities as class names.
This indicates that when overly long, repository-level information is provided as context, LLMs may hallucinate and undermine the accuracy of microservice decomposition.

% In Figure \ref{fig:exp1}, class "\textit{Item}" is regarded as a common class shared in both \textit{Order Service} and \textit{Catalog Service}, which is an unreasonable partition.
% Although "\textit{Item}" seems linguistically similar and contextually related to domain objects like "\textit{LineItem}","\textit{CartItem}" in \textit{Orde Service}, 
% if we look into the relevant code snippet, we can find that "\textit{Item}" represents the product item of suppliers, which should exclusively belong to the \textit{Catalog Service} domain, a different domain from \textit{Order Service}.

For gulimall application, the decomposition result contains three partitions reflecting different application domains: \textbf{Cart Service}, \textbf{Order Service}, and \textbf{Ware Service}.
In Figure \ref{fig:exp2}, LLM assigns the class ``\textit{OrderItemVo}'' as a common class shared between both \textbf{Cart Service} and \textbf{Order Service}.
% Although ``\textit{OrderItemVo}'' seems linguistically similar and contextually related to value objects like ``\textit{CartItemVo}'' in \textit{Cart Service}, 
% if we look into the relevant code snippet, we can find that ``\textit{OrderItemVo}'' represents the items in customers' orders, which is used in ordering and stock locking scenarios, and should belong to \textit{Order Service} and \textit{Ware Service}, instead of \textit{Cart Service}.
However, a closer inspection of the code reveals that ``\textit{OrderItemVo}'' models items within a customer order and is used in order submission and stock-locking workflows. 
Accordingly, it should be placed in \textbf{Order Service} and \textbf{Ware Service}, rather than in \textbf{Cart Service}.
This error suggests that the LLM may rely heavily on surface cues (e.g., similarity between ``\textit{OrderItemVo}'' and ``\textit{CartItemVo}'') while failing to fully incorporate usage context and call relationships.
% This indicates that LLM may lack insight in code context, without further inspection into the code elements.
% identify domain-specific classes as shared common classes.
% It also overlooks the microservice design principles, as this decomposition may lead to unnecessarily high coupling among different domains.
Consequently, LLM fails to adhere to microservice design principles, which leads to unnecessarily high coupling among different services.

% (1) Over-Reliance on Semantic Similarity:
% LLMs tend to classify based on semantic understanding, such as grouping classes with names containing "item" together, even though they belong to different domains in terms of functionality.

% (2) Hallucination on Code Structure:
% When both code and database schema are input, LLMs may mistakenly interpret database information as class names, interfering with accurate identification.

% (3) Inaccurate Common Classes Identification and Assignment:
% LLMs tend to identify domain-related classes as part of the common classes, or assign common classes to the service they perceive as using them most frequently, fail to fully recognize common classes, or cannot accurately distribute common classes to the appropriate services based on dependency relationships.

\textbf{Our Insights.}
From these observations, we derive three key insights:

% \textit{\#1 Combination of Traditional Code Analysis and LLM Semantic Capabilities:}
% To address issues arising from relying solely on global code analysis or LLMs, we combine both approaches. 
% We use lightweight class dependency analysis to extract critical dependency information, providing it as context to the LLM. 
% This allows the LLM to make comprehensive decisions by integrating its semantic understanding of code with domain knowledge of microservices.

\textit{\#1 Utilization of a multi-agent framework to divide complex tasks.}
We break down the intricate task of microservice decomposition into multiple subtasks, creating a workflow that is collaboratively executed by multiple agents.
This approach narrows the focus for each agent, allowing them to concentrate solely on their designated subtask and avoiding the burden of managing excessive context from a large, comprehensive task.
% During the domain aggregation phase, which requires a fine-grained view of each domain, we also dynamically generate the required number of agents based on domain needs, with each agent focusing exclusively on its assigned domain.

\textit{\#2 Customizing hierarchical context for better utilization of context window.}
Since directly providing raw application code and database information to agents is impractical, we preprocess and compress the original context before it is used by LLMs in their decomposition subtasks. 
Through extraction, summarization, and relationship analysis, we construct context at varying granularities, both at the application level and class level, and carefully select the scope of context to expose to each agent.

\textit{\#3 Specialized tools designed to assist further analysis.}
To aid with agents' understanding of the code, we design analytical tools that enable them to explore underlying context among classes and align with core microservice design principles.
% For example, the agent can call entropy calculation tool to facilitate the identification of common classes, and employ a verification tool to check class assignments by performing a propagation algorithm.
% \vspace{-0.175in}
\section{Methodology}
\label{sec:methodology}
\begin{figure*}[t]
  \centering
  \includegraphics[width=0.98\linewidth]{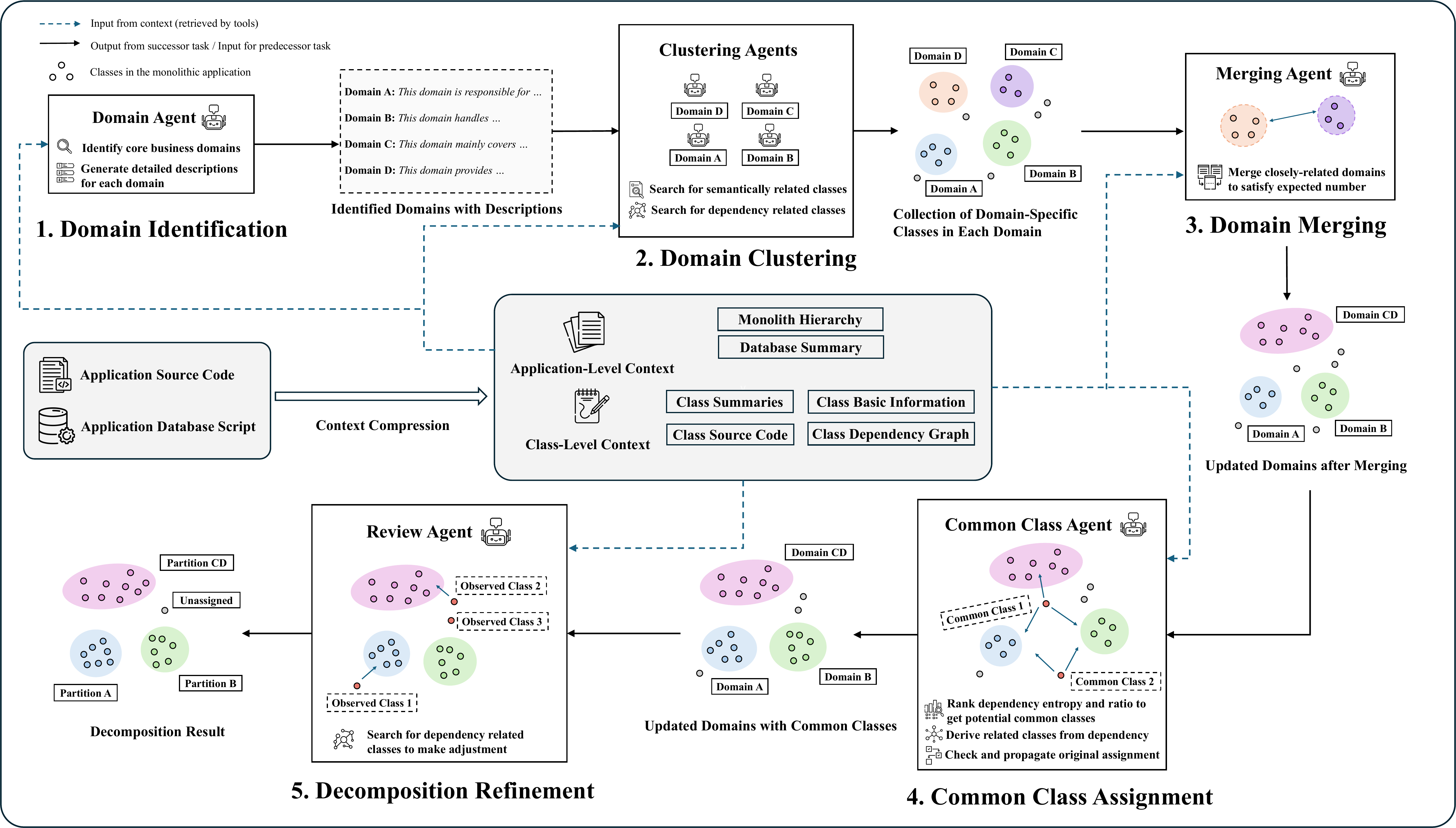}
  \caption{Overview of the multi-agent workflow of \tool. 
  % Here the five white boxes (with specific agent name, subtask names and indexes) depict each subtask. 
  % The content between every two boxes represent the output of the previous subtask, and meanwhile the input for the next subtask.
  % The two gray boxes in the middle refer to the relevant and available context provided for the agent(s) via tool calling.
  }
  \Description{}
  \label{fig:method_overview}
  % \vspace{-0.2in}
\end{figure*}
Figure \ref{fig:method_overview} illustrates the framework of \tool for microservice decomposition. 
To address the problems of overlong context length and limited context insights, we split the complex decomposition task into five subtasks, and assign specialized agent(s) to each subtask.
These agents collaborate to achieve the shared goal of effective microservice decomposition.

Our design of these five subtasks adheres to the principles of domain-driven design (DDD), which consists of two phases~\cite{vernon2013implementing}:
(1) \emph{strategic phase} that identifies business domains and establishes application domain boundaries;
(2) \emph{tactical phase} that further refines the structure and responsibilities within each bounded context.
In microservice architecture, each bounded context serves as a microservice candidate.
Accordingly, our first three agents implement the strategic phase (discovering and shaping bounded contexts), while the last two agents complete the tactical phase (refining the contents of each context).

% Based on this foundation, we decompose the task into five subtasks, each equipped with the corresponding agent(s).
Specifically, in the first subtask \textit{Domain Identification}, the \textbf{Domain Agent} is responsible for analyzing business logic and identifying candidate domains.
% Once the business domains are identified, the process moves to the \textit{Domain Clustering} subtask.
Then, in \textit{Domain Clustering}, the \textbf{Clustering Agents} are dynamically instantiated for each domain, where each agent collects and clusters domain-specific classes associated with its respective domain.
To avoid overly fragmented contexts and to meet the desired number of microservices, in \textit{Domain Merging}, the \textbf{Merging Agent} evaluates the identified domains and merges closely related ones into more cohesive bounded contexts, completing the strategic phase of DDD.
The remaining subtasks correspond to the tactical phase of DDD.
In \textit{Common Class Assignment}, the \textbf{Common Class Agent} identifies classes shared across domains and assigns them to the appropriate ones.
Finally, the process concludes with \textit{Decomposition Refinement}. 
The \textbf{Review Agent} examines remaining unassigned classes and decides whether to incorporate them into existing domains or leave them unassigned based on relevance and cohesion.
The resulting partitions of this subtask constitute the final microservice candidates.

Throughout this process, agents enhance their contextual understanding through tool invocation. 
As illustrated in Table \ref{tab:tool}, each agent has access to a distinct set of tools. 
There are two types of agent tools: the retrieval tools and the specialized tools.
The retrieval tools provide raw contextual information directly, while the specialized tools perform in-depth analysis, integrating microservice design principles. 
The context available to agents is organized into two levels of granularity, including the application-level context and the class-level context.
In the rest of this section, we will introduce the design of context management and multi-agent workflow in detail. 

\begin{table*}[t]
  \caption{Tool type, descriptions, and agents accessible.}
  % \vspace{-10pt}
  % \small
  \label{tab:tool}
  \centering
    \resizebox{0.95\textwidth}{!}{

  \begin{tabular}{p{0.18\textwidth}p{0.32\textwidth}p{0.75\textwidth}p{0.2\textwidth}} 
    \toprule
    \textbf{Tool Type} & \textbf{Tool} & \textbf{Tool Descriptions} & \textbf{Agents} \\ 
    \midrule
    \multirow{7}{*}{\makecell[l]{Retrieval\\Tools}} & get\_class\_hierarchy & Get monolith hierarchy. & \makecell[l]{All Agents} \\ 
     & get\_database\_summary & Get database summary for the application. & \\ 
    \cmidrule(lr){2-4}
     & get\_all\_class\_summaries & Get summaries for all classes. & \makecell[l]{Domain Agent} \\ 
     & get\_class\_summary & Get summary for single class input. & \\ 
     & get\_class\_basic\_info & Get basic information for input class. & \\ 
    \cmidrule(lr){2-4}
     & get\_class\_relation\_and\_code & Get relationship and source code for input class. & \makecell[l]{All Except Domain Agent} \\ 
     & search\_file & Search for a term in a file and return matching lines. & \\ 
    \midrule
    \multirow{5}{*}{\makecell[l]{Specialized\\Tools}} & codebase\_semantic\_search & Search for class summaries that are semantically related to a natural language query. & \makecell[l]{Clustering Agents} \\ 
    \cmidrule(lr){2-4}
     & get\_related\_class\_list & For a list of classes, find all classes they depend on and all classes that depend on them, and return the unique union. & Clustering Agents, \newline Review Agent \\ 
    \cmidrule(lr){2-4}
     & rank\_dependency\_entropy\_and\_ratio & Calculates both incoming and outgoing dependency entropy and ratio for all classes. & \makecell[l]{Common Class Agent} \\ 
     & get\_more\_potential\_common\_classes & Takes a list of potential common classes and returns all their dependents and dependencies that could be additional potential common classes. & \\ 
     & assign\_common\_class\_list & Takes a dict of common classes with their domain assignments and relationship types, and return a propagated assignment. & \\ 
    % \midrule
    \bottomrule
  \end{tabular}
}
% \vspace{-10pt}
  % \caption{Tool type, descriptions and agents accessible}
\end{table*}

\subsection{Context Compression}
% \sw{include this section into the overview figure or add a new figure for it.}
\label{subsec:compression}
As discussed in Section~\ref{sec:introduction} and Section~\ref{sec:background_motivation}, directly providing repository-level code to the agents would be excessively verbose and could be a burden to their memory.
To ensure the agents receive context at an appropriate level of granularity, we compress the context in advance from two perspectives: application-level and class-level.
% information for different subtasks in microservice decomposition, which involves extraction, analysis, or summarization.
% \tbd{Need to reconsider the order of these context?}
% \red{Application-level context is accessible to all agents, offering a global and preliminary view of the codebase and data structure. 
% In contrast, class-level context is divided further: the Domain Agent is provided with a brief class-level sketch to understand the functionalities of the entire monolithic application and to identify all business domains. 
% For the subsequent four agents, detailed class-level information is made available to support precise decomposition. }

\subsubsection{Application-Level Context}
We extract application-level context and provide it to all agents, offering a global and preliminary view of the code and database structure.
% , since they offer a global and preliminary context of the codebase and data structure for the decomposition process.
It includes:

\textit{(1) Monolith Hierarchy:}
% To provide agents with a comprehensive understanding of all packages and classes within the monolithic application, 
We extract the package and class lists directly from the source code and organize them in a hierarchical format. 
% In this hierarchy, the first level represents packages, while the second level represents classes within those packages. 
This hierarchical structure offers a concise, top-level overview that helps agents infer potential functionalities based on package divisions.

\textit{(2) Database Summary:}
Database definitions in monolithic applications can vary from SQL schema to non-SQL document.
% To provide a unified and more LLM-friendly\cite{?} structure, 
Therefore, we preprocess the database files by requiring an LLM to provide metadata summaries based on the database file to provide a unified and more LLM-friendly~\cite{ni2024next} structure.
% This preprocessing step ensures a more LLM-friendly data structure, facilitating efficient retrieval and interpretation by the agents \cite{?}.

\subsubsection{Class-Level Context}
% While application-level context provides a global view for agents, class-level context is crucial for accessing context in classes or across classes.
% The class-level context is further divided to class-level sketch (containing class summaries and class basic information) and class-level details (containing class dependency graph and source code).
% They each provide a different view of the codebase information.
% Here we introduce the content of them.
While application-level context offers a global perspective for agents, class-level context is essential for accessing detailed information within individual classes or across multiple classes. Class-level context is further categorized into two components: \textit{class-level sketch}, which includes class summaries and basic information,
%\sw{if you don't use these two terms (i.e., sketch and detail)in the following context (e.g., fig.2, each component), do not use them. Just say ``from two perspectives" is great.} 
%\sw{need examples here to illustrate what the basic info is like. use e.g., or i.e.,} 
and \textit{class-level details}, which encompass the class dependency graph and source code. 
% Each component provides a unique perspective on the codebase, complementing to offer a comprehensive understanding. 
The Domain Agent is provided with a brief class-level sketch to understand the functionalities of the entire monolithic application and to identify all business domains. 
For the subsequent four agents, detailed class-level information is made available to support precise decomposition.
Below, we introduce the contents of each component in detail.
% local information. \todo{add class-level sketch and details}

% For the Domain Agent that needs to grasp the functionalities of the whole monolithic application and identify all business domains, it would be unrealistic to delve into the source code of each class, which causes an overlong context. 
% Hence, we provide brief class-level sketch for it while minimizing information loss, containing class summaries and class basic information.

% However, unlike Domain Agents which utilizes a comprehensive overview of the application and classes, the subsequent four agents need access to class-level details for accurate decomposition.
% Therefore, apart from the class source code, we also provide the class dependency graph for them to retrieve and analyze.

% Here is the introduction of the processed class-level context:

\textit{(1) Class Summaries: }
% To distill the source code of each class while maintaining its expressiveness, 
We utilize an LLM to generate a compact and informative summary for each class.
These summaries include: (a) the main purpose of the class, (b) its key methods and the functionalities of the methods,
%\sw{a and b is too abstract. What do you mean by purpose? and how to identify if a method is key. pls avoid using pronouns}
and (c) any additional significant information identified by the LLM. 
This distills the source code of each class while maintaining its expressiveness.
% This approach effectively compresses the code context while minimizing information loss, ensuring that the essential details are preserved.

\textit{(2) Class Basic Information: }
% To provide a structured yet brief representation of each class without overwhelming the agents with excessive detail, 
We sketch the class by extracting key elements of each class and wrapping them in a JSON format.
The key elements include the class name, fields, method signatures, annotations, and dependencies.
% This approach delivers a clear abstract of each class, enabling agents to understand its core attributes efficiently.
This context delivers a clear abstract of each class, enabling agents to understand its core attributes efficiently.

\textit{(3) Class Dependency Graph: }
% Dependency Analysis has been proved to be a practical method to track syntactic and semantic relations among code elements\cite{compilers, codeplan}. 
% Since our focus is on class-level decomposition, we specifically analyze class dependency relationships within the monolithic application. 
We track syntactic and semantic relations among code elements by applying class dependency analysis.
% Our analysis identifies 11 distinct class relationship types, as detailed in Table \ref{tab:relations}, and defines common patterns for each type to ensure compatibility with popular Java web application frameworks \cite{static analysis of java enterprise applications}.
For syntactic analysis, we utilize JavaParser \cite{javaparserJavaParserHome}, a lightweight static analysis tool, to generate an Abstract Syntax Tree (AST). 
%\sw{the usgae of JavaParser can be removed to the implementation (sub)section. Here we just say what we do in the logical level. like, We construct AST, xxx, for each class. Further, we apply semantic derivation by using CHA, xxx. Need a brief intro to AST and CHA.}
For semantic derivation, we employ Class Hierarchy Analysis (CHA) \cite{dean1995optimization}. 
% In the resulting class dependency graph, each outgoing edge from node A to node B represents that class A depends on class B.
% while each incoming edge indicates that class A is a dependency for another class.
This information is also serialized into JSON format.

\subsection{Multi-Agent Workflow}

% \todo{reconsider the structure}
% In this subsection, we will introduce each subtask in the multi-agent workflow. For each stage, We first introduce the role and target for the corresponding agent, and then we analyze the specific context and the toolkit design.

In this subsection,
%\sw{sometimes we don't use it,so that we don't care if it's a subsection or section. Just use sth. like In order to / To xxx, we design/ propose a five-stage multi-agent workflow.},
we present an overview of each subtask within the multi-agent workflow. For each stage, we begin by outlining the role and objectives of the corresponding agent. Next, we delve into the specific context required for the task and detail the design of the associated toolkit.

% Among these, the context for the agent is selected from the different granularities of context mentioned in the previous subsection, based on the specific role and target for each agent.
% Meanwhile, we also design tools for the agent to call, as shown in Table \ref{tab:tool}.
% The tools are divided into two categories -- 
% carefully assign the necessary tools to each agent to strengthen the contextual awareness and microservice 
% As Table \ref{tab:tool} demonstrates, based on the specific tasks each agent is in charge of, we carefully assign the necessary tools to each agent to strengthen the contextual insight and 
% These tools are divided into two

\subsubsection{\textbf{Domain Identification}}
% \sw{it would be better and clearer if we could specify each stage's input and output}

At this stage, the goal for \textbf{Domain Agent} is to develop a comprehensive understanding of the entire monolithic application and identify core application domains.
Given the desired number of target microservices $n$,
the agent identifies a set of at least $n$ domains and generates a detailed description for each domain.
% the output goal for this agent is to identify at least $n$ domains and generate detailed descriptions for each domain.
% \sw{not sure what the term means}

\textbf{Context Analysis.}
This stage involves conducting a high-level overview of the monolithic application without delving into the implementation details. 
Therefore, we provide the agent with \textit{class summaries} and \textit{basic class information}, rather than details related to source code or class dependencies.

\textbf{Toolkit Design.}
To gather application-level information, we introduce the \textit{get\_all\_class\_summaries} tool, which allows the agent to obtain an outline of the entire application.
Given that applications typically consist of numerous classes, 
we additionally provide class-level tools that allow the agent to selectively access information about individual classes.
These include the \textit{get\_class\_summary} and \textit{get\_class\_basic\_info} tools.
% and considering that class summaries are already a condensed form of context, 
% we also offer class-level tools to consolidate memory and prevent memory lapse as aforementioned. 
% \sw{the 1st and 2nd reasons do not show a logical relationship with the choice. and it's a bit hard to understand their relationships through this sentence.}
% , which enable the agent to review specific classes in detail.

% \vspace{-0.09in}
\subsubsection{\textbf{Domain Clustering and Domain Merging}}
\label{subsec:domain clustering merging}
The objective of these two subtasks is to derive a qualified set of domains with the classified domain-specific classes.
The agents cluster domain-specific classes for the domains identified in the preceding subtask and merge closely related domains when necessary.

During the \textbf{domain clustering} stage, the identified domains and their corresponding descriptions obtained from the first subtask serve as inputs for dynamically instantiated \textbf{Clustering Agents}. 
Each domain is assigned to a dedicated Clustering Agent, ensuring that the number of agents corresponds to the number of domains identified.
%\sw{is the number consistency so important and will be re-mentioned in the following context? if no, we can consider removing the sentence.}
The task description for each agent is also dynamically generated, incorporating the specific domain's objectives along with descriptions of all other domains, as illustrated in Figure~\ref{fig:method_overview}. 
This enables each agent to be aware of domain boundaries and can accurately cluster classes within its assigned domain.
The output of each agent is a list of classes associated with its respective domain, and the clustering results across all domains are concatenated for use in subsequent subtasks.

The \textbf{domain merging} subtask is optional and is invoked if the number of identified domains exceeds the desired number of microservices.
%\sw{better indicate the desired number of ms is necessary in the background section}. 
In such cases, the \textbf{Merging Agent} is employed. 
This agent is provided with the class collections from all domains and is tasked with merging certain domains to achieve the target number of microservices while preserving domain coherence.

\textbf{Context Analysis.}
For both subtasks, agents are supposed to determine the specific functionalities and interactions of classes, necessitating a more fine-grained context regarding the \textit{source code} and \textit{class dependencies}. 
For the Clustering Agent, however, the challenge lies in narrowing the scope of analysis to a relevant subset of classes for a particular domain. 
We therefore design specialized tools to assist this agent.

\textbf{Toolkit Design.}
To facilitate access to code details and class dependencies, we provide two retrieval tools: \textit{get\_class\_relation\_and\_code} and \textit{search\_file}.
These tools are utilized by both the Clustering Agents and the Merging Agent.
As discussed earlier, narrowing the scope of domain clustering is critical for effective microservice decomposition. 
To achieve this, we design specialized tools tailored for the Clustering Agent. 
Drawing inspiration from the cohesion characteristics of microservices~\cite{lewis2014microservices,thones2015microservices}, we leverage both semantic search and dependency analysis to refine the scope of clustering. 
The following tools support this process:

\textit{(i) codebase\_semantic\_search}:
Motivated by RAG technique~\cite{lewis2020retrieval}, this tool employs a pre-trained sentence transformer~\cite{huggingfaceSentencetransformersallMiniLML6v2Hugging} to generate embeddings for each class based on its summary. 
The Clustering Agent can input a natural language query (e.g. domain description) into the tool, and this tool calculates the semantic similarity between the query and the class summaruy of each class.
The top-K most semantically similar class summaries are returned as the result. 
This enables the agent to identify candidate classes that are semantically aligned with the target domain, providing an initial focus for further analysis.

\textit{(ii) get\_related\_class\_list}:
While the \textit{codebase\_semantic\_search} tool identifies classes based on semantic similarity, this tool provides a complementary perspective by focusing on class dependencies.
With a list of class nodes as inputs, this tool extracts all the reachable class nodes that have a direct dependency/dependent relationship with any node in this list.
By integrating dependency analysis with semantic similarity, this tool enhances the agent's ability to understand the structural and functional relationships within the codebase, thereby improving the precision of domain clustering.

% \vspace{-0.09in}
\subsubsection{\textbf{Common Class Assignment}}
In this subtask, the target for \textbf{Common Class Agent} is to identify common classes and assign them to the corresponding domains.

% Common classes here refer to the classes that should be partitioned into more than one microservice partition, which are not directly related to domain business logic (and hence not identified in precedent subtasks), but provide common service for more than one doamin. 
% Motivated by the previous study\cite{ASE}, we identify two types of common classes based on the roles they play in applications: 

Common classes refer to classes that need to be allocated into multiple microservice candidates. 
% These classes are not directly tied to domain-specific business logic but provide shared services across domains. 
Unlike domain classes that implement domain-specific business logic rules, common classes provide cross-domain support or cross-cutting capabilities.
%\sw{too abstract, will be better if we can list some examples.}

% Motivated by the previous research \cite{wang2024microservice}, we identify two types of common classes based on their roles:\sw{each type needs a short, highly condensed name so that readers can quickly know it's reasonable to have these two types identified.}
% (1) Common classes that are commonly used across multiple domains as utility or helper classes. 
% % These classes typically provide shared functionality or services that are not domain-specific. 
% These classes offer shared functionality across multiple domains and are not domain-specific.
% (2) Common classes that act as dependents of multiple domains, serving as coordinators or intermediaries among domains.\sw{need some examples like the first type (e.g, utility/helper)}
% \tbd{Need a code figure for these?}

Motivated by the previous research \cite{wang2024microservice}, we distinguish two types of common classes based on the direction of cross-domain dependencies:
(1) \emph{Multi-domain depended-on common classes}: classes that are depended on by multiple domains.
Typical examples include reusable utilities (e.g., utility/helper classes) and shared data contracts (e.g., DTO/VO/TO models).
(2) \emph{Multi-domain dependent common classes}: classes that depend on multiple domains.
These classes typically serve as coordinators or mediators across domains, such as facades or orchestration components.

The agent first identifies these two types of common classes, then detects the domains that depend on or are dependent on by each common class, and finally assigns the class to the corresponding domains based on these dependency relations.

% The agent first identifies these common classes, and then determines the domains that need them\sw{too colloquial, use ``invoke/dependent on" will be better}, and assigns them accordingly.
% assigns them respectively to the corresponding domains that depend on them or serve as dependents of them either directly or indirectly.

\textbf{Context Analysis.}
% At this stage, context related to \textit{source code} and \textit{class dependencies} are critical to Common Class Agent to distinguish whether a class is common class or not.
% However, the difficulty would be, given the enormous number of classes in the monolithic applications, how to locate potential common classes with less missing rate.
% Specialized tools should be designed to give a deeper analysis towards the classes relationship.
At this stage, understanding the context of \textit{source code} and \textit{class dependencies} is crucial for the Common Class Agent to determine whether a class should be shared across multiple domains. 
Given the large number of classes in monolithic applications, locating potential common classes with minimal oversight is challenging. 
Therefore, specialized tools are necessary for in-depth analysis of class relationships.
% \sw{using more linking words will improve the readability and fluency of the paper}

\textbf{Toolkit Design.}
% To enable the agent to explore the details of classes, we still provide the retrieval tools \texttt{search\_file} and \texttt{get\_class\_relation\_and\_code} at this stage.
% To tackle with the challenge of circling the range of possible common classes, we design three specialized tools to provide statistical information:
To enable the agent's exploration of class details, we provide retrieval tools like \textit{get\_class\_relation\_and\_code} and \textit{search\_file}.
% \texttt{search\_file} and \texttt{get\_class\_relation\_and\_code}. 
To further assist the agent in identifying the potential common classes accurately, we also design three specialized tools:

\textit{(i) rank\_dependency\_entropy\_and\_ratio}:
To facilitate the discovery of the two types of common classes mentioned above, with the inspiration from the metrics measuring purity in microservices\cite{kalia2020mono2micro,wang2024microservice}, we propose to use Dependency Entropy, utilizing Shannon entropy\cite{shannon1948mathematical} to measure the ``domain purity" of classes.
Based on the two dependency directions, Dependency Entropy contains incoming and outgoing dependency entropy.
% \sw{as the metric we finally adopted is Dependency Entropy, need to highlight it here. Like, we propose to use Dependency Entropy, a variant of Shannon entropy, to measure xxx. And then include a brief intro of it, indicating that it has incoming and outgoing dependency entropies.}
An intuition is that classes that are commonly used by multiple domains, such as utility classes, should have high incoming dependency entropy, indicating low domain purity.
% various domains depend on them.
And for the common classes that depend on multiple domains, like coordinator classes, should have high outgoing dependency entropy.
% \sw{depend on other domains and are dependent on by other domains is more correct? a bit confused}

Hence, we utilize Dependency Entropy to measure how diverse a class's dependencies are distributed across different business domains. 
% It is calculated in two ways: incoming and outgoing. 
% Here we take incoming dependency entropy as an example to illustrate the definition.
% It is calculated for incoming and outgoing dependencies, respectively.
We compute this metric separately for incoming and outgoing dependencies.
% Regarding the incoming dependency entropy $H_{in}(C)$ for class $C$, the tool first identifies $\mathcal{D} = \{D_1, D_2, \ldots, D_k\}$, where $\mathcal{D}$ is the set of $k$ unique business domains $D_n$ that depend on class $C$, i.e. have at least an incoming edge, pointing from a class in this domain, towards the target class $C$.\sw{hard to understand.}
For the incoming dependency entropy $H_{in}(C)$ of a class $C$, the tool first collects the set of business domains
$\mathcal{D}=\{D_1, D_2, \ldots, D_k\}$,
where each domain $D_n$ is included if it contains at least one class that depends on $C$. In the dependency graph, this means there exists at least one edge from a class in $D_n$ to $C$. Thus, $k$ denotes the number of distinct domains that have incoming dependencies to $C$.
The incoming dependency entropy is defined as:
$H_{in}(C) = - \sum_{i=1}^{k} P(D_i) \log_2(P(D_i))$,
where $P(D_i) = \frac{\text{Number of classes from domain } D_i \text{ that } C \text{ depends on}}{\text{Total number of classes that depends on C}}$.
The tool ranks all the classes and returns the classes with the top-{K} highest entropy.
Also, it attaches the calculation result of the incoming dependency ratio and outgoing dependency ratio, e.g., $Ratio_{in}(C) = \frac{N_{in}(C)}{N_{out}(C)}$, providing additional context by giving the fraction of incoming and outgoing dependencies.
These ratios indicate the dominant dependency direction (incoming vs. outgoing), which helps characterize common classes when bidirectional dependencies exist.
% \sw{need to add the reason why we need to offer this additional context.}

% This tool helps filter out part of the domain-specific classes that have active interactions with multiple domains, since, due to the high cohesion characteristics of microservices, the domain purity should be high for these types of classes, leading to a low entropy. \sw{too long to understand. and, it seems like the domain purity is explained here, it should be explained where it is first used.}

This metric helps the agent filter out domain-specific classes that interact with multiple domains, which are not the common classes we aim to identify.
Under the microservice principle of high cohesion, such classes are expected to exhibit high domain purity, meaning that most of their dependencies stay within a single domain rather than being spread across domains. 
As a result, their dependency distribution is concentrated, and their dependency entropy remains low.

\textit{(ii) get\_more\_potential\_common\_classes}:
% Although tool \texttt{rank\_dependency\_entropy\_and\_ratio} helps uncover part of the potential common classes, to see whether there are more common classes, by receiving an input of class list, this tool get the lists of all the related classes that have direct edges on the dependency graph. This tool functions similar as \texttt{get\_related\_class\_list}, to promote the discovery of more related classes, except it also outputs a notice to remind that these are just possible candidates for the agent to take into consideration.
While the previous tool identifies some potential common classes, this tool explores further by retrieving lists of related classes with direct edges in the dependency graph, promoting the discovery of additional candidates.

\textit{(iii) assign\_common\_class\_list}:
With the assistance of tools (i)-(ii), the agent can discover adequate common classes candidates. By leveraging the retrieval tools to look into the code details and dependencies, it can make an initial decision on the common class assignment. 
However, the complexity of the assignment grows when there are multiple common classes with both incoming and outgoing dependency relationships. The complexity makes it difficult for the agent to give a complete and accurate assignment directly.
For example, if class $A,B,C$ are common classes identified by the agent, with dependency $A\rightarrow B \rightarrow C$, if class $A$ is assigned to a partition, then class $B,C$ should also be assigned since $A$ depends on them.
This propagation of assignment is non-trivial and can be easily neglected by agents.
% \sw{it's hard to understand why the example shows difficulty for LLMs. Need more complex examples or give reasons why LLMs will easily neglect these dependencies.Also, we fail to mention how LLM handles the example case.(need to add)}
% Hence, the assignment should be propagated through the common classes dependency.
% \tbd{Need a figure to illustrate this?}

In order to fully integrate dependency analysis with the code understanding ability of LLM, and to derive a more reliable assignment result, we propose a propagation algorithm to validate and update the current assignment.
Take incoming dependency as an example.
% As depicted in Algorithm \ref{alg:assign_common}, the tool first constructs an incoming dependency graph for the input common classes.
% Then it conducts a topological sort for this graph based on the dependencies.
% After that, for each topologically sorted class, we obtain its dependents and the assigned domains, and the domains are propagated based on the topological order, which then updates the current assignment.\todo{Modify the descriptions to be easier to understand; Add one sentence to sum up the benefit of the tool.}
As shown in Algorithm \ref{alg:assign_common}, the tool begins by constructing an incoming dependency graph for the input common classes (Line 2). 
Next, it performs a topological sort on this graph to establish the order of dependencies (Line 3). 
Following the sorting, the tool iterates through each class in the topological order, retrieves its dependents and their assigned domains (Line 4-5), and propagates the domains accordingly (Line 6-7). 
This propagation algorithm ensures that the current assignment is updated systematically based on the dependency structure.

\begin{algorithm}[tbp]
% \small
\footnotesize 
% \scriptsize
\SetAlCapSkip{2pt}
\caption{Common Class Propagation Algorithm}
\label{alg:assign_common}

\DontPrintSemicolon
\SetKwFunction{FNormalize}{Normalize}
\SetKwFunction{FTSort}{TopologicalSort}
\SetKwProg{Fn}{Procedure}{}{}

\Fn{AssignCommonClass(\textit{original\_assignment})}{
    \textit{incoming\_graph} $\gets$ Build graph where edge ``A $\gets$ B'' means ``B depends on A''\;
    % \textit{outgoing\_graph} $\gets$  Build graph where edge "B $\gets$ A" means "A depends on B"\;
    \textit{incoming\_sorted} $\gets$ \FTSort{\textit{incoming\_graph}}\;
    % \textit{outgoing\_sorted} $\gets$ \FTSort{\textit{outgoing\_graph}}\;

    % \textit{updated\_assignments} $\gets$ copy of \textit{normalized\_assignments}\;

    \BlankLine
    
    % $incoming\_reversed \gets$ Reverse all edges in $incoming\_graph$\;
    % $outgoing\_reversed \gets$ Reverse all edges in $outgoing\_graph$\;

    \For{\textit{common\_class} $\in$ \textit{incoming\_sorted}}{
        % \uIf{\textit{common\_class}.\textit{rel\_type} == 'incoming'}{
        %     $dependents \gets$ Get dependents of \textit{common\_class} from $incoming\_graph$\;
        % }
        \textit{dependents} $\gets$ Get dependents of \textit{common\_class} from \textit{incoming\_graph}\;
        \For{\textit{dependent} $\in$ \textit{dependents}}{
            % \textit{
            % Propagate the domain assignment of dependents to the original assignment of common\_class};
            \textit{updated\_assignment} $\gets$ Propagate the domain assignment of \textit{dependent} to \textit{common\_class}\;
        }
    }
    
    \KwRet{\textit{updated\_assignment}}\;
}
\end{algorithm}

% \input{algorithm/algo}
% \vspace{-0.09in}
\subsubsection{\textbf{Decomposition Refinement}}
% This is the final stage of the workflow, where the \textbf{Review Agent} is primarily tasked with examining any classes that remain unassigned and determining whether they should be incorporated into a domain. 
% Additionally, the agent can also make other necessary adjustments to optimize the decomposition. \sw{better to say the task of the Review Agent is to make necessary adjustments to optimize, including xxx and xx.}
% The output for this subtask will be the final partitions (i.e. microservice candidates) for the whole decomposition task.
% Hence, this subtask provides the last opportunity for the agent to review and refine the partitioning.\sw{The sentence shows little information.}
This is the final stage of the workflow. The \textbf{Review Agent} takes as input the current partitions produced by the preceding subtasks, together with the list of classes that remain unassigned, and is tasked with making necessary adjustments to improve the overall quality of the decomposition.
Its review process includes (i) inspecting classes that remain unassigned and deciding whether (and where) they should be incorporated into any partition; (ii) revisiting previously assigned classes to resolve inconsistencies.
The output of this subtask is the final set of partitions (i.e., microservice candidates) for the entire decomposition task. 
As the last step, it serves as a global check that consolidates earlier decisions and corrects residual issues before producing the final decomposition.

\textbf{Context Analysis.}
In this subtask, the agent focuses on context related to \textit{source code} and \textit{class dependencies}, as it explores the details of each unassigned class or any class requiring adjustment.

\textbf{Toolkit Design.}
Here we reuse the practical tools \textit{search\_file}, \textit{get\_class\_relation\_and\_code}, and \textit{get\_related\_class\_list} to enable the agent to acquire detailed information about the classes.
These tools are introduced in section \ref{subsec:domain clustering merging}.
% \sw{consider citing the corresponding section of the tool introduction, so that readers can quickly review tools' design and functionality.}

The complete prompt templates, including system prompts 
and task-specific user prompts, are provided in our artifacts.

\section{Evaluation}

We evaluate \tool by answering the following research questions (RQs):

\begin{itemize}[leftmargin=*]
    \item RQ1: How effective is \tool in microservice decomposition? 
    \item RQ2: How effective is \tool in assigning common classes to their microservice partitions?
    \item RQ3: How does each component contribute to \tool?
    \item RQ4: How does \tool perform with different LLMs?
\end{itemize}

\subsection{Experimental Setup}
\label{exp:setup}
\subsubsection{Benchmark Applications.}
Our benchmark consists of 10 Java applications, covering 49 services in total.
This benchmark relates to various business topics, and meanwhile includes the most common Java web frameworks~\cite{antoniadis2020static}.
%\sw{the sentence is less informative, consider removing it}
Among these ten applications, five are commonly used in previous works~\cite{wang2024microservice,nitin2022cargo,kalia2020mono2micro}:
% (1) Spring-Petclinic~\cite{githubGitHubSpringprojectsspringpetclinic}: a Spring Boot web application developed which implements a management system for pet clinic; (2) PartsUnlimitedMRP~\cite{githubGitHubMicrosoftPartsUnlimitedMRP}: a Spring Boot application on manufacturing resource planning; (3) 7ep-demo~\cite{githubGitHub7epdemo}: a Java EE web application developed to demonstrate software practices; (4) JPetStore~\cite{githubGitHubMybatisjpetstore6}: an online pet store application built with Spring, MyBatis and Stripes; (5) AcmeAir~\cite{githubGitHubBlueperfacmeairmonolithicjava}: a Jakarta EE web application designed for an airline company, where users can book tickets and manage their flights.
Spring-Petclinic~\cite{githubGitHubSpringprojectsspringpetclinic}, PartsUnlimitedMRP~\cite{githubGitHubMicrosoftPartsUnlimitedMRP}, 
7ep-demo~\cite{githubGitHub7epdemo}, JPetStore~\cite{githubGitHubMybatisjpetstore6}, and
AcmeAir~\cite{githubGitHubBlueperfacmeairmonolithicjava}.
The microservice versions for these monolithic applications can be publicly accessed~\cite{githubGitHubSpringpetclinicspringpetclinicmicroservices,githubGitHubMicrosoftPartsUnlimitedMRPmicro,githubBLUEPERF,wang2024microservice}.
Apart from these five applications, we additionally select five larger microservice repositories~\cite{youlai,passjava,zlt,goodskill,gulimall}, 
which satisfy the criteria: (1) the number of classes is more than 50; (2) stars on GitHub are more than 2k; (3) the line of code is larger than 5k; (4) the number of microservices is no less than 5.
Table~\ref{tab:benchmark} shows the details of our selected benchmark applications, including number of classes for decomposition, LOC (Line of Code), and number of microservice partitions.
% We construct the monolithic version of these applications with the assistance of LLM tools under human supervision and review.
% For these five applications with only microservice versions publicly available, we construct the corresponding monoliths with the assistance of LLM tools. 
% Each monolith is first validated to pass compilation, functional tests and E2E tests.
% Then these monoliths were independently reviewed by three authors to ensure no business logic is lost and no microservice-specific artifacts remain.

For the latter five applications with only microservice versions publicly available, we construct the corresponding monoliths following a two-stage protocol.
(1) \emph{Construction and validation}: two authors (each with at least three years of software development experience) independently used LLM-assisted tools to merge the microservice modules into monolithic applications and recorded intermediate documentation. Each constructed monolith was validated to pass compilation, functional tests, and end-to-end tests.
(2) \emph{Independent review}: a third author (with six years of software development experience) reviewed the intermediate documentation, independently inspected each monolith, and retested the monoliths.
Any issue identified by any author was discussed and resolved jointly before the monolith was accepted.

\begin{table}[tbp]
% \vspace{-0.1in}
\small
  \caption{Breakdown of our benchmark applications.}
  \label{tab:benchmark}
  \centering
  % \vspace{-0.1in}
  \resizebox{0.4\textwidth}{!}{
  \begin{tabular}{lccc}
    \toprule
    \textbf{Repository} & \textbf{\# of Classes} & \textbf{LOC} & \textbf{\# of Microservices}\\
    \midrule
    JPetstore & 24 & 1,409 & 3 \\
    Spring-Petclinic & 23 & 752 & 4 \\
    PartsUnlimitedMRP & 53  & 4,407 & 5\\
    7ep-demo & 47 & 2,326 & 4 \\
    Acme Air & 32 & 2,048 & 4 \\
    youlai-mall & 352 & 14,127 & 6 \\
    PassJava & 73 & 10,086 & 6 \\
    microservices-platform & 103 & 12,166 & 5 \\
    goodsKill & 143 & 7,856 & 5 \\
    gulimall & 379  & 20,714 & 8 \\
  \bottomrule
\end{tabular}
}
% \vspace{-13pt}
\end{table}

% \vspace{-0.08in}
\subsubsection{Baselines.}
% Our baselines include Mono2Micro~\cite{kalia2020mono2micro}, Data-Centric~\cite{romani2022towards}, CARGO~\cite{nitin2022cargo}, MOSAIC~\cite{filippone2023monolithic}, and an LLM baseline.
Our baselines include Mono2Micro~\cite{kalia2020mono2micro}, CARGO~\cite{nitin2022cargo}, MonoEmbed~\cite{sellami2026monoembed}, MOSAIC~\cite{filippone2023monolithic} and an LLM baseline.
% \sw{need some brief intros to these tools if possible.}
Among these, CARGO and MOSAIC are method-level decomposition tools, and the rest are all class-level decomposition tools.
% For the non-LLM approaches, part of the applications' decomposition results are run by the original tool authors with their recent versions and have been released in the study~\cite{wang2024microservice}.
% To ensure consistency and fairness, for the remaning results of MOSAIC and Data-Centric, we approach the authors and ask them to run the same version of tools;
% for Mono2Micro and CARGO developed by commercials, we run by ourselves with the most recent released code or package\cite{ibmMono2MicroOverview,githubGitHubIBMcodenetminervacargo}.
For the LLM baseline, we provide the original application code, database scripts, and the target number of microservices as input, and prompt the model in a zero-shot setting to generate decomposition results, thereby avoiding bias from few-shot examples~\cite{xu2022alleviating}.
% \todo{truncation?}

% \vspace{-0.08in}
\subsubsection{Models.}
We select three widely used LLM models: Deepseek-V3.2~\cite{liu2025deepseek}, GPT-5.2~\cite{openaiIntroducingGPT52}, and Claude Sonnet 4.5~\cite{anthropicClaudeSonnet} to conduct our experiment, due to their accessibility and stability.
For all models, we set the temperature to 0 to get a more deterministic result.
% We select three widely used LLM models: Deepseek-V3~\cite{liu2024deepseek}, GPT-4.1~\cite{openaiIntroducingGPT41}, and Claude Sonnet 4~\cite{anthropicClaudeSonnet} to conduct our experiment, due to their accessibility and stability.
% The detailed versions are Deepseek-V3-0324, gpt-4.1-2025-04-14, and claude-sonnet-4-20250514.
% The detailed versions are Deepseek-V3-0324, gpt-4.1-2025-04-14, and claude-sonnet-4-20250514.

% \vspace{-0.08in}
\subsubsection{Metrics.}
We use architectural metrics and similarity metrics to evaluate the overall decomposition results. 
Besides, we also verify the effects of correctly assigning common classes to further demonstrate the practicability of our decomposition results.

\textbf{Architectural Metrics.}
We adopt three widely used metrics from prior work to measure how well the decomposition follows microservice design principles:

(1) \textit{Code Modularity (CMod)}~\cite{wang2024microservice} measures the cohesion and coupling of the decomposition on the static call graph.
For each partition, it is computed as the ratio of intra-partition edge weights to the sum of intra-partition and inter-partition edge weights.
We calculate the modularity for each partition separately and compute the mean across all partitions. 
Higher values indicate better modularity.

(2) \textit{Cyclic (In-)Dependence (CiD)}~\cite{wang2024microservice} evaluates the proportion of partition pairs that are free from cyclic call dependencies. 
Higher values indicate fewer undesirable cyclic dependencies.

(3) \textit{Business Context Purity (BCP)}~\cite{kalia2020mono2micro, nitin2022cargo, wang2024microservice} measures how concentrated the business use cases are within each partition using entropy. 
Higher values indicate that each partition better aligns with the Single Responsibility Principle~\cite{singlePrinciple}.

% \textit{Domain Independence (DI)}~\cite{wang2024microservice,lewis2014microservices}:\todo{change to BCP}
% This assesses the number of partitions involved in managing each business use case\sw{the term is abstract and hard to understand}, reflecting how well the decompositions align with the business-domain-focused design principle of microservice architectures. 
% This metric is calculated using the entropy of a set of methods recorded in an execution log for a specific business use case.
% % samples, which, in this context, are 
% The probability that a use case is implemented in a partition $p$ is $P'(p)=\frac{n_p}{n}$, where $n_p$ is the number of methods in partition $p$ that implement the use case, and $n$ is the total number of partitions implementing the use case. \sw{higher is better? pls specify it.}
% % A higher DI indicates better

% This metric is complementary to BCP, it measures how well the decomposition results follow the by-business-domain decomposition principle.
% Its calculation involves the partition entropy across the business domains, and the method is similar to BCP.

\textbf{Similarity Metrics.}
% ~\sw{the aforementioned similarity metrics do not show here. consider replace the current version with it to ensure text consistency.}
As revealed by the recent study~\cite{wang2024microservice}, although architectural metrics have been widely utilized for evaluation in previous works, they can only offer a partial and theoretical view of the decomposition quality, since high architectural metrics can also lead to low quality or impractical decomposition results.
Therefore, to evaluate how accurate and practical our decomposition results are, we compare them with the ground truth decomposition results (i.e., the corresponding microservice versions).

\textit{Weighted Similarity (WS)}:
This metric measures the accuracy of a produced decomposition $P$ compared with the reference $R$, refining the calculation in~\cite{wang2024microservice} to compute in both directions.
For each partition $p_i$ in $P$, we find the most overlapping partition $r_j$ in $R$ and calculate
$Sim(p_i,r_j)= \frac{|p_i \cap r_j|}{\max(|p_i|, |r_j|)}$.
The directional weighted similarity is
$\text{WS}_{P\to R} = \sum_i Sim(p_i,r_j) \times w_j$,
where $w_j$ is the proportion of $|r_j|$ relative to the sum of all matched partition sizes. 
$\text{WS}_{R\to P}$ is computed analogously.
The overall weighted similarity is the harmonic mean of both directions. 
A higher value indicates higher similarity to the ground truth.
To ensure classes that are not considered in the reference decomposition do not influence the results, we exclude those uncounted classes (usually w.r.t. underlying infrastructure) in the produced decomposition before calculation.

\textit{Cluster-to-cluster Coverage ($c2c_{cvg}$)}~\cite{lutellier2017measuring,wang2024microservice}:
This metric represents to what extent a produced decomposition overlaps with the ground truth.
It measures the proportion of partitions whose best match exceeds a similarity threshold percentage $th_{cvg}$.
Following the same bidirectional scheme as weighted similarity, we define
$\text{c2c}_{(P \to R)} = \frac{|\{p_i : Sim(p_i, r_j) \geq th_{cvg}\}|}{|P|}$,
compute $\text{c2c}_{(R \to P)}$ analogously, and take the harmonic mean. 
We report results at 50\% (some overlap), 75\% (moderate overlap), and 90\% (high overlap).
For example, with $th_{cvg}=90\%$, a higher value of $\text{c2c}_{cvg}$ indicates that there are more partitions in $P$ that are at least 90\% similar to the reference decomposition $R$.

\textbf{Common Class Assignment Metrics.}
To further demonstrate how well \tool generates practical decomposition results, we evaluate the effects of common classes identification and assignment from the service level.
% Common classes refer to the classes shared across multiple partitions in the ground truth decomposition.
For each service $s$ and common class $c$, the reference allocation is defined as: 
$s$ should contain $c$ if $s$ includes any class that depends on $c$. 
We compare this reference allocation against the actual decomposition result to compute True Positives (TP, correctly allocated), False Positives (FP, redundantly allocated), and False Negatives (FN, missing allocations). 
Then we calculate the Precision, Recall, and F1 based on these.
Higher precision indicates fewer unnecessarily allocated common classes, while higher recall indicates that the decomposition misses fewer common classes that services depend on.
% For common classes in ground truth that are classified into more than one partition, we calculate their  Accuracy, Recall, and F1 rate to measure how correctly the produced decompositions identify and assign them.\sw{in theory, we need some formulas here}

% \vspace{-0.14in}
\subsection{RQ1: Effectiveness of \tool in Microservice Decomposition}
\begin{figure}[h]
  \centering
  \includegraphics[width=\linewidth, trim=5 10 5 10, clip]{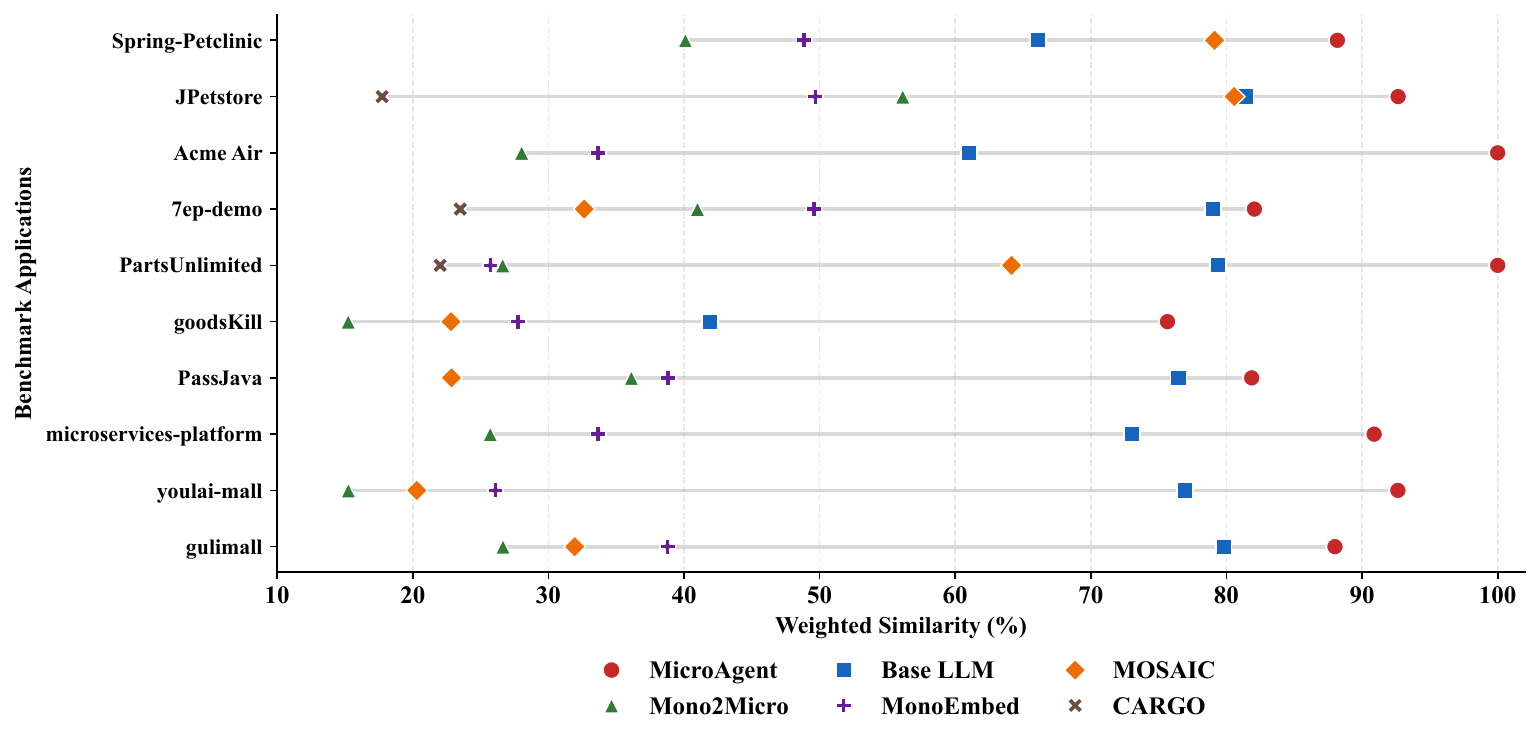}
  \caption{Per-application WS across methods.}
  \Description{}
  \label{fig:similarity_per_app}
  % \vspace{-0.15in}
\end{figure}

We evaluate the performance of our framework against the baselines with both architectural metrics and similarity metrics to answer RQ1.
The results are displayed in Table \ref{tab:RQ1} and Figure~\ref{fig:similarity_per_app}.
Both \tool and the base LLM model use Deepseek-V3.2 as backbone.

Among the approaches, four of six tools can produce results for all ten benchmark applications.
CARGO~\cite{nitin2022cargo} can only obtain results for three applications, and MOSAIC~\cite{filippone2023monolithic} fails to decompose two applications.
These limitations may stem from restricted framework compatibility, or from certain applications falling outside the tools' underlying assumptions.
% fails to obtain results for two applications, due to its incomplete static analysis~\cite{wang2024microservice} or incompatibility with the framework.
% MOSAIC~\cite{filippone2023monolithic} cannot work for decomposing AcmeAir since it makes the assumption that the target application should contain domain entities as classes, while this one is not.

% We compare the effectiveness of these methods with both architectural metrics and similarity metrics.

% \textbf{(1) Architectural Metrics:} 
In terms of \textbf{Architectural Metrics},
% Regarding coupling and cohesion, 
\tool achieves the highest scores in CMod, showing an 11.9\% improvement compared with the best baseline MOSAIC.
For cyclic dependency, \tool attains the highest percentage of 98.6\%, which is close to 100\%, indicating that our decomposition results are largely free of cyclic dependence.
For business context purity, \tool gains a lower score in BCP compared with MOSAIC.
By inspecting case studies, this is likely because \tool considers common classes shared across partitions, which reduces business context purity.
%\sw{any proposals to recover from it?}
% in two out of five applications, and also shows superior performance in the remaining three ones. 
% Among the four tools that successfully obtain results for all benchmark applications, \tool shows an average of 20.1\% improvement compared with the best baseline Mono2Micro.
% For cyclic dependency, \tool is the only tool that consistently achieves 100\%, indicating that our decomposition results are free of cyclic dependence for all benchmark applications.
% This metric also outperforms the base LLM model by 27.1\% on average, which is the best baseline.
% In terms of domain independence, our method surpasses the best baseline -- the LLM base model, among the four tools that can obtain complete results, with 26.1\% enhancement.
%
% \textbf{(2) Similarity Metrics:} 
For \textbf{Similarity Metrics},
% We also compare the results with the manually-produced ground truth microservice decomposition.
\tool exhibits significant improvement in the weighted similarity (i.e., accuracy) when compared with the ground truth decomposition.
It achieves an 89.2\% accuracy on average across the benchmark applications, which outperforms the best baseline, LLM base model, by 24.6\%.
Figure~\ref{fig:similarity_per_app} further demonstrates that \tool consistently surpasses other baseline methods across all the benchmark applications.
For the similarity coverage, our approach improves by 17.7\%, 36.9\%, and 409.6\% regarding the metrics of $c2c_{cvg}$ when the threshold is 50\%, 75\%, and 90\%, respectively.
% It achieves over 80\% accuracy for all benchmark applications, and 91.5\% on average, which outperforms the best baseline, LLM base model by 38.6\%.
% For the similarity coverage, our approach improves by 33.3\%, 112\%, and 425\% regarding the metrics of $c2c_{cvg}$ when the threshold is 50\%, 75\%, and 90\%, respectively.

\begin{table}[h]
\small
    \caption{Results of architectural metrics and similarity metrics for each approach. Bold indicates the best performance.}
    
% \vspace{-0.1in} 
    \label{tab:RQ1}
    \centering
\resizebox{\linewidth}{!}{
\begin{tabular}{c|lllllll}
\hline
\multicolumn{1}{c|}{\multirow{2}{*}{\textbf{Approach}}} & \multicolumn{3}{c|}{\textbf{Architectural Metrics}} & \multicolumn{4}{c}{\textbf{Similarity Metrics}} \\ \cline{2-8} 
                           &\textbf{CMod}   & \textbf{CiD}    & \multicolumn{1}{l|}{\textbf{BCP}} & \textbf{WS} & $\boldsymbol{th_{cvg}=50\%}$   & $\boldsymbol{th_{cvg}=75\%}$   & $\boldsymbol{th_{cvg}=90\%}$  \\ \hline

\multicolumn{1}{l|}\tool             &   \multicolumn{1}{c}{\textbf{93.0}}     & \multicolumn{1}{c}{\textbf{98.6}}  & \multicolumn{1}{l|}{34.6}   & \multicolumn{1}{c}{\textbf{89.2\%}} &                                    \multicolumn{1}{c}{\textbf{94.5\%}}  & \multicolumn{1}{c}{\textbf{75.7\%}} & \multicolumn{1}{c}{\textbf{53.0\%}}   \\
                           
                           \multicolumn{1}{l|}{Base LLM}  &    \multicolumn{1}{c}{82.9}    &  \multicolumn{1}{c}{89.6} & \multicolumn{1}{l|}{33.7}   & \multicolumn{1}{c}{71.6\%} & \multicolumn{1}{c}{80.3\%}  & \multicolumn{1}{c}{55.3\%}  & \multicolumn{1}{c}{10.4\%}    \\
                           
                           \multicolumn{1}{l|}{Mono2Micro}      &   \multicolumn{1}{c}{73.6}     &  \multicolumn{1}{c}{71.3}     & \multicolumn{1}{l|}{41.1}   & \multicolumn{1}{c}{31.1\%} & \multicolumn{1}{c}{14.4\%}   & \multicolumn{1}{c}{3.3\%}  & \multicolumn{1}{c}{0\%}    \\
                           
                           \multicolumn{1}{l|}{MonoEmbed}    &  \multicolumn{1}{c}{38.5}    &   \multicolumn{1}{c}{93.8}     & \multicolumn{1}{l|}{42.3}   & \multicolumn{1}{c}{37.3\%} & \multicolumn{1}{c}{22.0\%}  & \multicolumn{1}{c}{0\%}  & \multicolumn{1}{c}{0\%}   \\
                           
                           \multicolumn{1}{l|}{CARGO}           &  \multicolumn{1}{c}{42.6}    &   \multicolumn{1}{c}{58.9}    & \multicolumn{1}{l|}{26.6}   & \multicolumn{1}{c}{21.1\%} & \multicolumn{1}{c}{0\%}    & \multicolumn{1}{c}{0\%}   & \multicolumn{1}{c}{0\%}    \\
                           
                           \multicolumn{1}{l|}{MOSAIC}          &  \multicolumn{1}{c}{83.1}    &   \multicolumn{1}{c}{97.9}  & \multicolumn{1}{l|}{\textbf{45.5}}   & \multicolumn{1}{c}{44.3\%} & \multicolumn{1}{c}{36.9\%}  & \multicolumn{1}{c}{22.5\%} &  \multicolumn{1}{c}{8.3\%}   \\ \hline
\end{tabular}
}
% \vspace{-10pt}
\end{table}

\noindent
\colorbox{gray!20}{%
    \parbox{\dimexpr\linewidth-2\fboxsep}{
    % \small
    \textbf{Answer to RQ1:}
    \tool achieves satisfying results in both architectural metrics and similarity metrics.
    Our tool demonstrates a 24.6\% improvement in the weighted similarity, and a substantial 409.6\% improvement in $c2c_{cvg}$ ($th_{cvg}=90\%$). 
    This implies \tool produces decomposition results that are closer to the ground truth decomposition, which are of much higher quality.
    % Our tool demonstrates a 24.6\% improvement in the weighted similarity, and a substantial 409.6\% improvement in $c2c_{cvg}$ ($th_{cvg}=90\%$). 
    % This implies \tool produces decomposition results that are closer to the ground truth decomposition, which are of much higher quality.
    }%
}

\subsection{RQ2: Performance of \tool in Assigning Common Classes}
Assigning common classes is one of the most error-prone aspects of dependency management in microservice decomposition, especially in multi-domain scenarios where proper allocation is critical yet highly challenging. We compare how \tool and baseline tools perform in this complex task.

As shown in Table \ref{tab:RQ2}, the precision, recall, and F1 scores for each class-level decomposition approach are reported on average.
\tool shows the highest precision in identifying and assigning common classes, which is 3.4\% higher than the best baseline (Mono2Micro).
Moreover, it demonstrates remarkable improvement in both recall and F1, which are 52.7\% and 41.1\% higher than the best baseline (base LLM), respectively.
These indicate that \tool not only presents high recall in identifying necessary common classes that should be shared among partitions, 
but also precisely assigns the common classes to the corresponding microservice partitions.
% \tool shows high accuracy in identifying and assigning common classes, which is only 1.3\% lower than the best baseline -- base LLM.
% In contrast, it demonstrates remarkable improvement of recall and F1, which are 208.6\% and 108.5\% higher than the best baseline, respectively.
% These indicate that although some baseline approaches present high accuracy when assigning common classes to the corresponding microservice partitions, they miss a large proportion of necessary common classes in their decomposition.
% With the customized context and specialized tool design for our Common Class Agent in the framework, our approach can identify and assign all common classes, with an average recall rate of 100\%.

\begin{table}[h]
% \vspace{-10pt}
    \caption{Experiment results demonstrating the effects of common classes assignment. Bold indicates the best performance.}
    % \vspace{-0.1in}
    \label{tab:RQ2}
    \centering
    \resizebox{0.3\textwidth}{!}{
    \begin{tabular}{l|l|l|l}
    \toprule
    \textbf{Approach}      & \textbf{Precision}    &  \textbf{Recall}     &  \textbf{F1} \\ 
    \midrule
    \tool         &  \textbf{94.0\%}    &  \textbf{93.6\%}     &  \textbf{93.4\%} \\
    Base LLM      &  82.7\%    &  61.3\%    &  66.2\% \\
    Mono2Micro    &  90.9\%    &  31.5\%    &  45.2\% \\
    MonoEmbed     &  43.7\%    &  13.4\%    &  19.9\% \\
    \bottomrule
    \end{tabular}
    }
% \vspace{-10pt}
\end{table}

\noindent
\colorbox{gray!20}{%
    \parbox{\dimexpr\linewidth-2\fboxsep}{
    % \small
    \textbf{Answer to RQ2:}
    \tool showcases the highest precision and recall in common classes assignment.
    It notably improves the F1 score by 41.1\%, indicating the effectiveness of our designs in \tool for accurately identifying and properly distributing the common classes.
    }%
}

\subsection{RQ3: Ablation Study of \tool}
To evaluate the contribution of each component in \tool, we conduct an ablation study to assess their effectiveness.
We mainly focus on four important settings: (1) without the dependency analysis context, (2) without specialized tools for Clustering Agent, (3) without specialized tools for Common Class Agent, and (4) without multi-agent framework (single agent with all tools provided).
%\sw{do we have any reasons not to include other agents?}

Table \ref{tab:RQ3} demonstrates the ablation study results on average.
% Among these three settings, \tool surpasses 
When specialized tools are not provided for the Clustering Agent, it obtains the lowest score in BCP among these four settings.
Meanwhile, the overall quality and accuracy of the decomposition declines compared with \tool, as weighted similarity and $c2c_{cvg}$ ($th_{cvg}=90\%$) deteriorate by 6.8\% and 17.2\%, respectively.
% We can notice that under this setting, the F1 score also falls from 97.7\% to 82.7\%, approaching the degree of decrease when there are no common class tools.
% As we investigate the cases, we can find that this is mainly because unsatisfactory clustering results can propagate to subsequent phases. 
% For example, when the Clustering Agent fails to cluster certain classes for a domain in this process, then in Common Class Assignment process, the Common Class Agent may not be able to identify and assign the common classes correctly for this domain.
%

Under the setting without specialized tools for Common Class Agent, all types of metrics fall consistently, 
the decrease is pronounced for the $c2c_{cvg}$ when $th_{cvg}=90\%$. 
It drops from 53.0\% to 38.3\%, representing a 27.7\% decrease.
Meanwhile, the weighted similarity decreases from 89.2\% to 82.1\% with F1 score drops from 93.4\% to 92.0\%.
% the decrease is especially pronounced for the $c2c_{cvg}$ when $th_{cvg}=90\%$ -- it drops from 70.0\%  to 17.3\%, representing a significant 63.8\% decrease.
% Meanwhile, the weighted similarity decreases from 91.5\% to 78.8\%, and the F1 score drops from 97.8\% to 81.8\%.
This indicates that the specialized tools for Common Class Agent contribute to both the common class metric and the overall decomposition quality.%\sw{what does the metric stand for? like, higher common class metric is high, and leads to what? Making decomposition more practical or? }

Omitting the dependency context also results in worse performance.
There is a 10\% decrease in weighted similarity from 89.2\% to 80.2\%, and a significant 27.7\% drop in $c2c_{cvg}$ ($th_{cvg}=90\%$) from 53.0\% to 38.3\%.
Moreover, the F1 score also exhibits a decrease from 93.4\% to 91.5\%.
These all show that the dependency context is crucial for a more accurate decomposition.

We further evaluate a single-agent setting, where all tools are provided to one agent without the multi-agent workflow.
This setting yields the most significant degradation: weighted similarity drops by 13.8\% to 76.9\%,
and F1 decreases sharply from 93.4\% to 81.2\%.
This suggests that the multi-agent workflow is a critical component in \tool. 
By decomposing the task into focused subtasks, each agent can concentrate on a specific scope without being overwhelmed by overlong context, 
whereas a single agent struggles to coordinate the full decomposition even when provided with the same set of tools.

\begin{table}[ht]
\caption{Ablation study results of \tool. Bold indicates the best performance.}
% \vspace{-0.1in}
\label{tab:RQ3}
\centering
\resizebox{\linewidth}{!}{
\begin{tabular}{l|ccc|cccc|c}
\hline
\multicolumn{1}{c|}{\multirow{2}{*}{\textbf{Setting}}} & \multicolumn{3}{c|}{\textbf{Architecture Metrics}} & \multicolumn{4}{c|}{\textbf{Similarity Metrics}} & \textbf{Common Class} \\ \cline{2-9} 
\multicolumn{1}{c|}{}     & \textbf{CMod}        & \textbf{CiD}         & \textbf{BCP}         & \textbf{WS}  & $\boldsymbol{th_{cvg}=50\%}$       & $\boldsymbol{th_{cvg}=75\%}$     & $\boldsymbol{th_{cvg}=90\%}$      & \textbf{F1} \\ \hline
\tool                     & \textbf{93.0} & \textbf{98.6} & \textbf{34.6} & \textbf{89.2\%} & \textbf{94.5\%} & \textbf{75.7\%} & \textbf{53.0\%} & \textbf{93.4\%} \\
w/o clustering tools      & 91.9 & 93.6 & 29.4 & 83.1\% &  88.4\% & 75.0\% & 43.9\% & 92.2\% \\
w/o common class tools    & 92.2 &  93.3 & 32.6 & 82.1\% &  86.7\% & 71.8\% & 38.3\% & 92.0\% \\
w/o dependency context    & 92.0 & 94.6 & 32.9 & 80.2\% & 83.1\% & 65.6\% & 38.3\% & 91.5\% \\
single agent    & 91.1 & 93.6 & 33.4 & 76.9\% & 84.3\% & 59.9\% & 40.1\% & 81.2\% \\
\hline
\end{tabular}
}
% \vspace{-10pt}
\end{table}

\noindent
\colorbox{gray!20}{%
    \parbox{\dimexpr\linewidth-2\fboxsep}{
    % \small
    \textbf{Answer to RQ3:}
    % The ablation study shows the contribution of our design in context and tools for agents.
    % The accuracy of the decomposition result deteriorates consistently under all settings when one of the components is omitted. 
    The ablation study confirms the contribution of each component. 
    Among all settings, replacing the multi-agent workflow with a single agent causes the largest performance deterioration.
    The specialized tools and dependency context each further contribute to the overall decomposition quality.
    }%
}

\subsection{RQ4: Effects of Different LLMs}

We select three commonly used LLM models to study how different backbone models generalize on \tool. As introduced in section \ref{exp:setup}, they are Deepseek-V3.2, GPT-5.2, and Claude-Sonnet-4.5.
Table \ref{tab:RQ4} demonstrates the performance of \tool with these different base models.

As revealed by the similarity metrics, \tool can consistently improve the accuracy and quality of decomposition across all the base models.
\tool enhances the weighted similarity by 15.1\%-24.6\%, and increases $c2c_{cvg}$ ($th_{cvg}=90\%$) with an range of 185.5\% to 409.6\%.
In terms of identifying and assigning common classes,
\tool outperforms the original base model by 30.0\%-73.9\% on average.
Additionally, while the Claude base model surpasses the \tool in $c2c_{cvg}$ ($th_{cvg}=50\%$), \tool clearly outperforms it in weighted similarity, F1, and $c2c_{cvg}$ with $th_{cvg}=90\%$.
This suggests that Claude base model more often produces decompositions that are moderately aligned with the ground truth (above 50\% similarity), whereas \tool consistently delivers highly accurate results, achieving substantially more near-exact matches (above 90\% similarity) and better overall performance.
Also, we can notice that \tool obtains the greatest improvement in weighted similarity and $c2c_{cvg}$ ($th_{cvg}=90\%$) when applied to Deepseek-V3.2, and achieves the most enhancement in F1 for Claude-Sonnet-4.5 model.
% % \tbd{add a clarification?}
% Also, we can notice that DeppSeek base model performs worst in similarity and common class metrics, but achieves the best performance when equipped with \tool; while Claude base model performs the best among these three base LLMs, it showcases the least substantial improvement with \tool.

\begin{table}[tbp]
% \vspace{-10pt}
\caption{Generalization of different LLMs with \tool.}
% \sw{we can bold the best performance per LLM. It is always our tool.}} 
% \vspace{-0.1in}
\label{tab:RQ4}
\centering
\resizebox{\linewidth}{!}{
\begin{tabular}{l|l|ccc|cccc|c}
\hline
\multicolumn{1}{c|}{\multirow{2}{*}{\textbf{Model}}} & \multicolumn{1}{c|}{\multirow{2}{*}{\textbf{Approach}}} & \multicolumn{3}{c|}{\textbf{Architecture Metrics}} & \multicolumn{4}{c|}{\textbf{Similarity Metrics}} & \textbf{Common Class}    \\ \cline{3-10} 
\multicolumn{1}{c|}{}                       & \multicolumn{1}{c|}{}   & \textbf{CMod}          & \textbf{CiD}          & \textbf{BCP}         & \textbf{WS}   & $\boldsymbol{th_{cvg}=50\%}$  & $\boldsymbol{th_{cvg}=75\%}$  & $\boldsymbol{th_{cvg}=90\%}$  & \textbf{F1} \\ \hline

\multirow{2}{*}{Deepseek-V3.2}                & \tool                   & 93.0 & 98.6 & 34.6 & 89.2\% & 94.5\% & 75.7\% & 53.0\% & 93.4\% \\
                                            & Base LLM                & 82.9 &  89.6 & 33.7 & 71.6\% &  80.3\% & 55.3\% &  10.4\% & 66.2\% \\ \hline

\multirow{2}{*}{GPT-5.2}                    & \tool                   & 93.6 & 98.0 & 31.8 & 82.7\% & 90.9\% & 73.2\% & 47.4\% & 91.5\% \\
                                            & Base LLM                & 82.3 & 91.7 & 30.3 & 70.0\% &  84.1\% & 51.7\% &  16.6\% & 70.4\% \\ \hline

\multirow{2}{*}{Claude-Sonnet-4.5}            & \tool                   & 93.0 &  98.3 & 34.2 & 84.4\% &  86.3\% & 78.8\% & 54.0\% & 92.7\% \\
                                            & Base LLM                & 85.8 &  90.9 & 25.1 & 73.3\% &  93.2\% & 55.4\% & 16.3\% & 53.3\% \\  
\hline
\end{tabular}
}
% \vspace{-10pt}
\end{table}

\noindent
\colorbox{gray!20}{%
    \parbox{\dimexpr\linewidth-2\fboxsep}{
    % \small
    \textbf{Answer to RQ4:}
    \tool can consistently enhance the performance of the three LLM base models.
    It outperforms the base models in the weighted similarity by 15.1\%-24.6\%, and improves $c2c_{cvg}$ ($th_{cvg}=90\%$) by 185.5\%-409.6\%.
    This shows that \tool can generalize well with different backbones of LLM.
    }%
}

% \subsection{Case Study}\sw{this can be a section, apart from the evaluation section}
% \tbd{add a case study}
\section{Case Study}
% We conduct a case study on the mall application gulimall~\cite{gulimall} to illustrate how \tool performs microservice decomposition in practice.
% We trace four representative classes in the decomposition process: ``\textit{OrderItemVo}'', a shared data contract related to the order and warehouse domains; ``\textit{Query}'', ``\textit{SQLFilter}'', and ``\textit{RRException}'', a set of shared utility classes.

Recall from Section~\ref{sec:background_motivation} that when directly prompted with the repository-level context, the base LLM incorrectly assigned ``\textit{OrderItemVo}'' to \textbf{Cart Service} and \textbf{Order Service} based on surface-level name similarity (Figure~\ref{fig:exp2}).
Moreover, traditional tools often neglect common class assignment, resulting in misplaced or missing allocations for shared classes.
In this case study, we trace how \tool handles the same application gulimall~\cite{gulimall}, focusing on four representative classes: ``\textit{OrderItemVo}'', ``\textit{Query}'', ``\textit{SQLFilter}'', and ``\textit{RRException}''.

\textbf{Domain identification and clustering.}
Instead of providing the full repository-level information that can cause hallucination and information overload, Domain Agent operates on compressed class summaries and correctly identifies eight business domains, including \textbf{Order Service}, \textbf{Warehouse Service}, and \textbf{Member Service}.
Each domain is then handled by a dedicated Clustering Agent with domain-tailored context, which leverages \textit{codebase\_semantic\_search} and \textit{get\_related\_class\_list} to collect domain-specific classes (e.g., ``\textit{OrderServiceImpl}'' $\to$ \textbf{Order Service}, ``\textit{WareSkuServiceImpl}'' $\to$ \textbf{Warehouse Service}).

% \begin{figure*}[t]
%   \centering
%   \includegraphics[width=0.7\linewidth]{figures/case_study1.pdf}
%   \caption{
%   }
%   \Description{}
%   \label{fig:case_study1}
% \end{figure*}

% \begin{figure*}[t]
%   \centering
%   \includegraphics[width=0.7\linewidth]{figures/case_study2.pdf}
%   \caption{ 
%   }
%   \Description{}
%   \label{fig:case_study2}
% \end{figure*}

\begin{figure}[h]
  \centering

\includegraphics[width=0.98\linewidth]{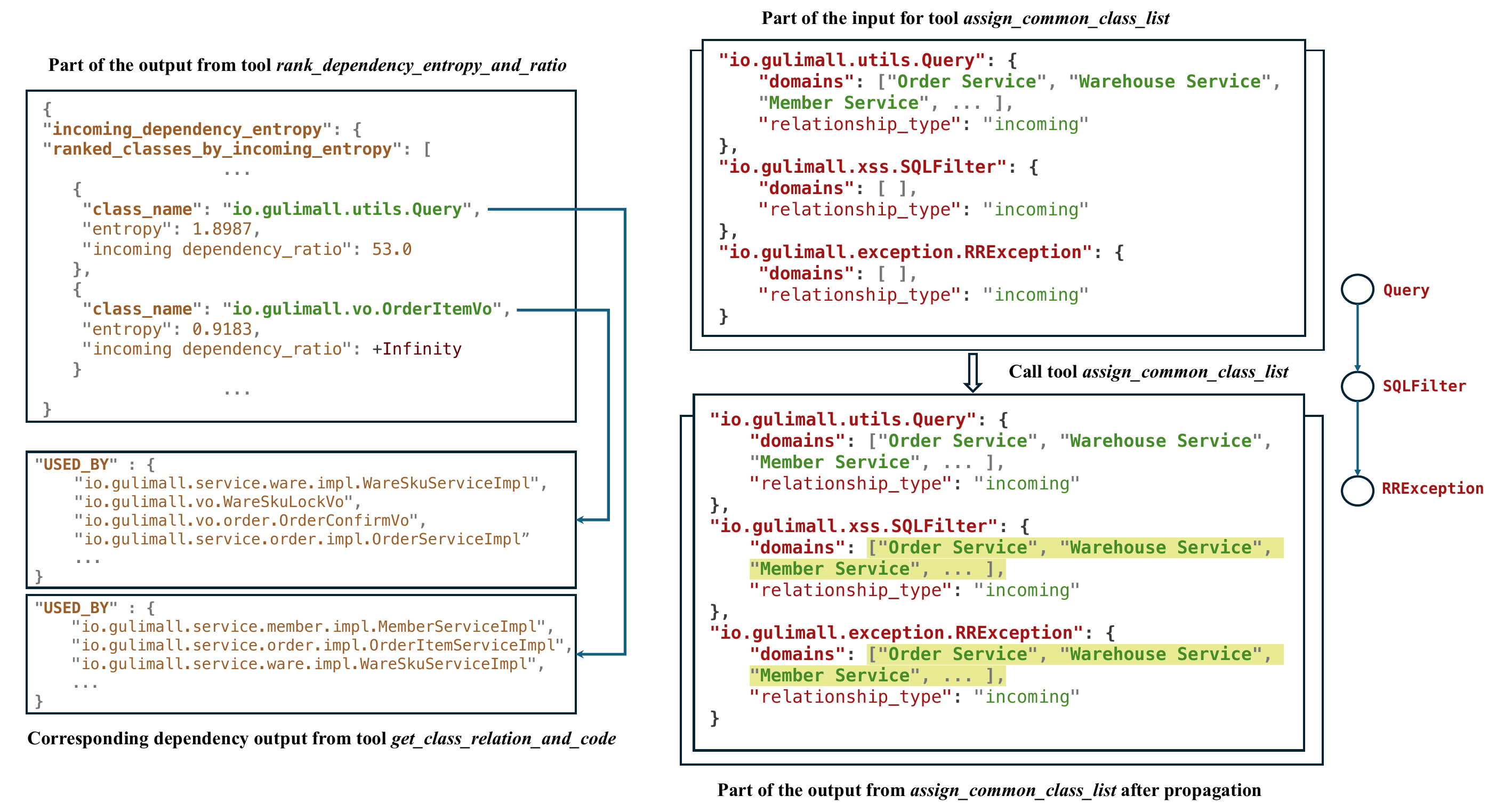}
  % \begin{subfigure}[t]{0.7\linewidth}
  %   \centering
  %   \includegraphics[width=\linewidth]{figures/case_study1.pdf}
  %   \caption{}
  %   \Description{}
  %   \label{fig:case_study1}
  % \end{subfigure}

  % \vspace{0.8em}

  % \begin{subfigure}[t]{0.7\linewidth}
  %   \centering
  %   \includegraphics[width=\linewidth]{figures/case_study2.pdf}
  %   \caption{}
  %   \Description{}
  %   \label{fig:case_study2}
  % \end{subfigure}

  \caption{Demonstration of case study in gulimall application.}
  \Description{}
  \label{fig:case_study}
  % \vspace{-10pt}
\end{figure}

\textbf{Common class identification and assignment.}
As shown on the left of Figure~\ref{fig:case_study}, the Common Class Agent invokes tool \textit{rank\_dependency\_entropy\_and\_ratio}, which flags both ``\textit{OrderItemVo}'' and ``\textit{Query}'' with high incoming dependency entropy, indicating they are depended on by multiple domains.
Unlike base LLMs, which rely on the name similarity between ``\textit{OrderItemVo}'' and ``\textit{CartItemVo}'' and assign the former class to \textbf{Cart Service}, the agent calls \textit{get\_class\_relation\_and\_code} to inspect actual usage context (shown in Figure~\ref{fig:exp2}): 
``\textit{OrderItemVo}'' is created during order submission in \textbf{Order Service} and consumed by ``\textit{WareSkuServiceImpl}'' for inventory locking in \textbf{Warehouse Service}.
Based on this code-level evidence, the agent correctly assigns ``\textit{OrderItemVo}'' to \textbf{Order Service} and \textbf{Warehouse Service}, matching the ground truth.
Similarly, ``\textit{Query}'', a pagination utility referenced across domains, is assigned to five domains including \textbf{Order Service}, \textbf{Warehouse Service}, and \textbf{Member Service}.

\textbf{Propagation for transitive dependencies.}
``\textit{Query}'' depends on ``\textit{SQLFilter}'', which in turn depends on ``\textit{RRException}''.
Neither class has direct cross-domain references, making them invisible to baseline approaches that only track first-order dependencies.
The agent invokes \textit{assign\_common\_class\_list}, which propagates domain assignments along the dependency chain (Algorithm~\ref{alg:assign_common}): the five domains of ``\textit{Query}'' are propagated first to ``\textit{SQLFilter}'', then to ``\textit{RRException}'' (Figure~\ref{fig:case_study}, right).
The final assignment for all four traced classes matches the ground truth, illustrating how the combination of multi-granularity context, code-level dependency analysis, and entropy-guided tools produces practical decomposition results.
% \section{Discussion}

% \vspace{-0.12in}
\section{Threats to Validity}

\textbf{Internal Validity.}
(1) Uncertainty of LLM Outputs:
The inherent non-determinism of LLMs may lead to variability in their outputs, which can impact the reproducibility of our experimental results. 
To mitigate this issue, we set the temperature parameter to 0 to ensure deterministic outputs as much as possible.
(2) Assumption of Known Microservice Partition Count:
We assume the target number of microservices is provided as known information in the input, following the assumptions of previous works~\cite{nitin2022cargo, kalia2020mono2micro, assunccao2021multi} in microservice decomposition.
However, in some real-world decomposition scenarios, the target microservice count may not always be predetermined and may be adjusted based on actual requirements.
The topic of finding the optimal number of microservices is out of the discussion scope in this paper.

\noindent \textbf{Construct Validity.}
(1) Data Leakage Risk:
The benchmark applications are open-source projects that may have appeared in the LLMs' training corpora.
Since \tool and the base LLM share the same underlying model, the relative comparisons among LLM-based approaches remain valid, though the absolute performance might be equally inflated.
(2) LLM-assisted Monolith Construction:
Five monoliths in our benchmark are reconstructed from their microservice repositories with LLM assistance, which may raise concerns that the constructed monoliths favor LLM-based approaches.
We consider this bias limited, as the construction mainly involves mechanical transformations (e.g., unifying build configurations and replacing inter-service calls with local invocations) that rarely introduce ambiguous design decisions.
Moreover, the ground-truth decomposition is determined by the original human-authored microservice repository, independent of the assistant model.

\noindent \textbf{External Validity.}
In our evaluation, we focus on applications in Java.
This is consistent with previous works in microservice decomposition,  since Java has been a dominant language for web or enterprise application development~\cite{antoniadis2020static}.
For generalizability in different programming languages, our framework design is language-agnostic and can also be applied to other languages. 
% \vspace{-0.12in}
\section{Related Work}

% Most of existing works follow a similar paradigm for microservice decomposition.
% %\sw{need some citing of the previous works}
% They first analyze a monolithic application and encode information to construct a relation graph, and then utilize clustering algorithms or community detection techniques to optimize a set of architectural metrics.
Most existing microservice decomposition studies follow a common paradigm: they construct a relation graph from a monolith and apply clustering or community detection to optimize architectural metrics.
% \sw{what does architectural metrics mean? we need to directly indicate that they can perform well or sth. else, not to mention an abstract metric here}.
These works can be categorized based on the information they utilize as follows.

The most common heuristics utilize code analysis to build application graphs, representing code elements as nodes and their relationships (e.g., function calls or control flow) as edges~\cite{nitin2022cargo,kalia2020mono2micro,assunccao2021multi,filippone2023monolithic,liu2022log2ms}.
% This is the most common category, as it includes abundant works.
These works typically utilize static analysis~\cite{nitin2022cargo,filippone2023monolithic}, dynamic analysis~\cite{kalia2020mono2micro,liu2022log2ms}, or combining both~\cite{assunccao2021multi,sellami2022combining} to construct a method-level or class-level dependency graph before clustering.
% For example, CARGO\cite{} utilizes context-sensitive static analysis to build a system dependency graph and adopts a partitioning algorithm based on label propagation.
% Their techniques are usually built upon static analysis tools like SOOT, WALA, or DOOP, which may ignore certain features of some common Java web frameworks\cite {elephants}, resulting in sparse coverage of the applications and cannot scale well for larger cases.
% ; Mono2Micro\cite{} creates use cases and traces the runtime relationship with a code instrumenter, and then applies clustering on execution traces.
However, due to the intrinsic limitations of static analysis, works that solely use static analysis typically have sparse coverage of code elements and fail to scale for larger cases.
% Moreover, the off-the-shelf static analysis tools they choose can only support certain frameworks in Java applications, and may have implementation issues when applied to popular frameworks like Spring.
% On the other hand, dynamic analysis requires a large collection of use cases, and can be intrusive (e.g. code instrumentation) to the original application running.
Moreover, dynamic analysis often requires a large collection of use cases, which can be difficult to achieve.
Due to these drawbacks, the graphs obtained by the program analysis techniques are usually sparse and incomplete, leading to poor clustering effects.

To address the limitations of program analysis, semantic analysis is often integrated to learn graph representations before clustering.
% Most previous works utilize existing techniques such as Term Frequency or embedding models, which only capture semantic features like class name similarity, without further information about the code context as well as the business logic.
Previous works~\cite{sellami2022combining,mazlami2017extraction,al2021microservice,trabelsi2023legacy} rely on Term Frequency or embedding models, which capture surface-level semantics like class name similarity but overlook code context and business logic.
% There have also been works~\cite{desai2021graph} that encode class embeddings via learning, which requires training for each application.
% Recent works~\cite{sellami2026monoembed,alsayedmicrodec} also leverage LLMs to embed code elements into a graph, while MicroDec~\cite{alsayedmicrodec} directly utilizes LLMs as tokenizers, MonoEmbed~\cite{sellami2026monoembed} fine-tunes on existing LLMs before using them, which requires substantial data collection.
Other approaches require per-application training for class embeddings~\cite{desai2021graph}. Similarly, recent LLM-based methods either treat LLMs as simple tokenizers~\cite{alsayedmicrodec} or demand data-intensive fine-tuning~\cite{sellami2026monoembed} before deployment.
% Previous works employ neural network training\cite{CoGCN} or model fine-tuning to generate embeddings for code\cite{contrast}, which often requires substantial data collection.
Additionally, some methods leverage external information for graph representation, including database schemas~\cite{romani2022towards} and version history~\cite{mazlami2017extraction}.
However, existing approaches generally lack sufficient understanding of 
the overall business logic, resulting in poor microservice decomposition performance in real-world scenarios.
In contrast, \tool leverages tailored multi-granularity context and decomposition-oriented tools to achieve deep comprehension of both codebase and business logic.

% \vspace{-0.13in}
\section{Conclusions}
We propose \tool, a context-augmented multi-agent framework for microservice decomposition. 
To address the inefficiencies of manual methods and existing automatic tools, 
\tool divides decomposition into five subtasks, equips specialized agents with tailored multi-granularity context, and provides decomposition-oriented tools aligned with microservice principles.
Across benchmark datasets, \tool achieves 89.2\% average decomposition accuracy, improving over the best baseline by 24.6\%, and reaches a 93.4\% F1 score on common-class identification and assignment, surpassing baselines by 41.1\%.
These results show that \tool can produce accurate and practical microservice partitions.
% \vspace{-0.13in}
\section*{Data-Availability Statement}
The artifact for our paper is publicly available at~\cite{su_2026_19250602}.
% \url{https://doi.org/10.5281/zenodo.19250602}.

\begin{acks}
This work is supported by the Research Grants Council Grant for Theme-based Research Scheme Project (RGC Ref. No. T43-513/23-N).
\end{acks}

\bibliographystyle{ACM-Reference-Format}
\bibliography{software}

\end{document}